\documentclass[a4paper,11pt]{article}
\usepackage{jheppub} 
\usepackage{graphicx}
\usepackage{subcaption}
\usepackage{multirow}
\usepackage{graphicx}
\title{\boldmath Probing Charm Yukawa Coupling through $ch$ Associated Production at the Hadron Colliders}

\author[a]{Nuoyu Dong,}
\emailAdd{2022111030003@stu.hznu.edu.cn}
\author[a]{Hongsheng Hou,}
\emailAdd{hshou@hznu.edu.cn}
\author[a]{Zhuoni Qian,}
\emailAdd{zhuoniqian@hznu.edu.cn}
\author[a]{Bowen Wang,}
\emailAdd{bowenw@hznu.edu.cn}
\author[a]{Pei Xu,}
\emailAdd{2022111030014@stu.hznu.edu.cn}
\author[a]{Qingjun Xu}
\emailAdd{xuqingjun@hznu.edu.cn}

\affiliation[a]{School of Physics, Hangzhou Normal University, Hangzhou, Zhejiang 311121, China}

\abstract{
At present, the study of the charm-quark Yukawa coupling at the Large Hadron Collider mainly focuses on the Higgs decay processes. Such signal suffers from overwhelming QCD background and derives its sensitivity primarily from the $Vh$ associated production channel. In addition, sensitivity to a possible CP phase in charm-quark Yukawa at the hadron collider is not discussed. We investigate the charm-Higgs associated production signal, that contains a potentially detectable interference term between $cch$ Yukawa coupling mediated diagrams and $ggh$ coupling mediated diagram. Such interference term is sensitive to the relative CP phase between contributing diagrams. High dimensional kinematic information are exploited by machine learning techniques to separate the different contribution, and sensitivity on the coupling is derived. Assuming a real $\kappa_c$ modification framework, 1$\sigma$ bound of $-5.6 < \kappa_c < 5.6$ (HL-LHC) and $-1.51 < \kappa_c < 1.62$ (FCC) are achieved. When allowing for CP-phase in the charm Yukawa, a combined 1$\sigma$ bound of $0.32<|\kappa_c|<1.69$, $-77^\circ < \alpha < 77^\circ$ (HL-LHC) and $0.70<|\kappa_c|<1.29$, $-55^\circ < \alpha < 55^\circ$ (FCC) can be achieved on the magnitude and CP phase of the coupling respectively.
}

\begin{document}
\maketitle
\flushbottom

\section{Introduction}
\label{sec:intro}

Since the discovery of the Higgs boson by ATLAS~\cite{ATLAS:2012yve} and CMS~\cite{CMS:2012qbp}, determining the properties of the Higgs boson has been a crucial task at the Large Hadron Collider (LHC). Current constraints from Run 2 LHC data show that coupling strengths between all third-generation fermions and the Higgs boson deviate by no more than 20\% from the Standard Model (SM) predictions~\cite{ATLAS:2018kot,CMS:2018nsn,ATLAS:2018ynr,CMS:2018vqh,CMS:2018nak,ATLAS:2018mme}.
At the High Luminosity (HL)-LHC, it is expected that the precision of the third-generation Yukawa couplings will be measured to within 5\%~\cite{MammenAbraham:2021ssc,Grojean:2020ech,Cepeda:2019klc}.
It is worth noting that in general a complex CP-phase can be introduced in Yukawa couplings in the presence of new physics contribution. Currently, LHC data put bounds on the top and tau Yukawa coupling CP-phase of: $(11^{+52}_{-73})^\circ$, $(9^{+16}_{-16})^\circ$~\cite{ATLAS:2024fkg} respectively. At the HL-LHC, constraints are expected to reach $|\alpha_t| < 36^\circ$~\cite{Goncalves:2021dcu}, $|\alpha_b| < 23.4^\circ$~\cite{Grojean:2020ech} for top and bottom, and $|\alpha_\tau| < 8^\circ$~\cite{Harnik:2013aja,Cepeda:2019klc} for tau at $2\sigma$ confidence level.
Measurements of Yukawa couplings in the second generation remain challenging, among which the muon Yukawa observation is the most promising. The current observed upper limit on $h\to\mu\mu$ decay rate is 2.2 times the SM prediction~\cite{ATLAS:2020fzp}, signifying an $\mathcal{O}$(1) muon Yukawa coupling measurement in the near future. The allowed range of the charm Yukawa coupling is measured to be 8.5 times that of the SM at 95\% confidence level from ATLAS~\cite{ATLAS:2022ers} or $1.1<\kappa_c<5.5$ from CMS~\cite{CMS:2022psv}. 

With the accumulation of data at the HL-LHC, extensive exploration on the charm Yukawa coupling are expected. 
Following the $\kappa$-framework, constraints on possible rescaling of SM couplings are described through $\kappa_i = g_i/g^{SM}_i$. We compile here a representative list of proposals and prospected bounds on charm Yukawa coupling in Table~\ref{tab:current-bounds}, under HL-LHC dataset. 
Among them, the Higgs rare decay channel $H \to J/\psi \gamma$ offers a small but clean signal in detecting the charm Yukawa coupling at the LHC~\cite{Bodwin:2013gca,Kagan:2014ila,Konig:2015qat}. A current constraint of $\kappa_c/\kappa_\gamma \in (-133,175)$~\cite{CMS:2022fsq,ATLAS:2022rej} is already performed using this signal with 139 fb$^{-1}$ data at $\sqrt{s} = 13$ TeV. A future reach of $0.32<\kappa_c/\kappa_\gamma<1.53$ assuming a $10\%$ precision on the decay branching is targeted for example at FCC~\cite{Konig:2015qat}. Similarly, $h \to c \bar{c} + J/\psi$ decay provides another exclusive hadron decay channel to further constrain the coupling~\cite{Han:2022kts}.
Precision measurement of the differential distribution of the Higgs from inclusive Higgs production are claimed to constrain the light quark Yukawa coupling as well~\cite{Bishara:2016jga, Soreq:2016rae}. With current Run 2 data, $\kappa_c \in (-4.46, 4.81)$ is achieved~\cite{ATLAS:2022qef, CMS:2024xsa}, and an order one reach is expected at the HL-LHC.  
Other proposals include the $ h \to c \bar{c} \gamma $ decay where the electroweak loop contribution is non-negligible~\cite{Han:2017yhy}, and associated production of charm quark with di-boson processes $ pp \to V V c X $~\cite{Vignaroli:2022fqh}. All could offer contribution to probing the charm Yukawa coupling, awaiting real simulation to be performed, and realistic bounds to be achieved. Note that the $h \to J/\psi + \gamma$ and the Higgs differential signal give asymmetric bounds on a real $\kappa_c$ modification, which come from sizable interference term that could be extrapolated to meaningful bounds on possible CP-phase of the coupling at HL-LHC and FCC.
The $Vh$ ($h \to c \bar{c}$) channel allows for a direct probe of the Higgs decay to a pair of charm jets~\cite{Perez:2015aoa, Perez:2015lra, Delaunay:2013pja, Carpenter:2016mwd,CMS:2022psv}. A prospects of $|\kappa_c|<2.5$ is projected from current Run 2 data and offers a realistic estimate for HL-LHC reach~\cite{ATL-PHYS-PUB-2018-016}. 
Other processes involving the Higgs decay to a pair of charm jet also contribute, such as the Higgs pair production $hh \to c \bar{c} \gamma \gamma$~\cite{Alasfar:2019pmn}, and the vector boson fusion (VBF) $ p p \to q q h \gamma(h \to c \bar c)$~\cite{Carlson:2021tes}. All the production modes collectively improve bounds on the decay branching and therefore improve bounds on magnitude of charm Yukawa coupling.
Much more precise determination of the branching ratio are expected to be obtained at future facilities such as the Large Hadron electron Collider (LHeC)~\cite{Li:2019xwd}, future lepton collider such as the Circular electron positron collider (CEPC)~\cite{An:2018dwb}, or a high energy muon collider~\cite{Forslund:2022xjq}. The prospected bounds are all summarized as in Table \ref{tab:current-bounds}.

\begin{table}
  \centering
  \begin{tabular}{|c|c|c|}
    \hline
    channel & Machine &  Bound   \\
    \hline
    Higgs differential & HL-LHC  & $-0.6 < \kappa_c < 3.0(2\sigma)$~\cite{Bishara:2016jga}  \\
    $h \to c \bar c \gamma$ &  HL-LHC   & $|\kappa_c| < 6.3(2\sigma)$~\cite{Han:2017yhy}\\
    $hh \to c \bar c\gamma\gamma$ & HL-LHC & $ -4.8 < \kappa_c < 4.6(2\sigma)$~\cite{Alasfar:2019pmn}\\
    $p p \to q q h \gamma(h \to c \bar c)$ & HL-LHC & $|\kappa_c| < 13(2\sigma)$~\cite{Carlson:2021tes} \\
    $pp \to VVcj$ &  HL-LHC   & $ -2.4 < \kappa_c < 1.77(2\sigma)$~\cite{Vignaroli:2022fqh}\\
    $h \to J/\psi +c \bar c  $  & HL-LHC  & $|\kappa_c| < 2.4(2\sigma)$~\cite{Han:2022kts} \\
    $pp\to c h $ & HL-LHC & $|\kappa_c| < 2.6(2\sigma)$~\cite{Brivio:2015fxa}\\
    $p p\to Z h(h\to c \bar c)$  & HL-LHC  & $|\kappa_c| < 2.5(2\sigma)$~\cite{ATL-PHYS-PUB-2018-016} \\
    $h \to J/\psi + \gamma$& FCC-hh  & $0.32 < \kappa_c/\kappa_\gamma < 1.53(2\sigma)$~\cite{Konig:2015qat}  \\
    $e^- p\to \nu_{e}hj(h\to c \bar c)$ & LHeC & $|\kappa_c| < 1.18(2\sigma)$~\cite{Li:2019xwd}  \\
    $e^- e^+ \to Zh(h\to c \bar c)$ & CEPC &  $0.98<\kappa_c < 1 .02(2\sigma)$~\cite{An:2018dwb}   \\ 
    $\mu^+ \mu^- \to \mu^+ \mu^-(\nu_{\mu} \bar{\nu_{\mu}})h(h\to c \bar c)$ & Muon Collider & $0.94< \kappa_c < 1.06(1\sigma)$~\cite{Forslund:2022xjq}   \\ 
    \hline
  \end{tabular}
  \caption{The table summarizes representative proposals and prospective constraints on the Charm-Yukawa coupling at the HL-LHC and FCC-hh, while the last three rows include constraints from related studies on future hadron-electron collider, electron-positron collider, and muon collider.}
  \label{tab:current-bounds}
\end{table}

The $c h$ associated production process
probes the coupling from the production side and tags the charm jet in the final state exclusively. An early analysis from Ref.~\cite{Brivio:2015fxa} estimates a sensitivity to the charm Yukawa coupling from the production rate as a function of the coupling. It however lacks a study in Higgs decay nor simulation with realistic effects. It also neglects contribution from the interference term which could become detectable with future data. To achieve an realistic sensitivity and understanding of the channel motivates the work of this study.
Building upon the foundation of the literature~\cite{Brivio:2015fxa}, we further conduct event simulation and analysis, provide a more realistic conclusions. Moreover, we consider a possible CP-odd component of the coupling and its probe through relevant collider signals.

The remaining content of the article proceeds as follows. In \autoref{sec:signal}, we introduce the chosen signal and all relevant backgrounds, provide the details for data simulation, and define a set of observable. 
In \autoref{sec:analysis}, we present event distributions of signal and background. We then analyze and categorize the simulated data with BDT and interpret the optimization with shapley values. In \autoref{sec:results}, we translate the categorization (confusion matrix) to constraints on the charm Yukawa coupling and possible CP-phase at the HL-LHC and FCC-hh. We combine complementary collider constraints as well as bounds electron dipole moment (EDM) measurement. Finally, in \autoref{sec:conclude}, we conclude. A derivation and the numerical dependence of the modified $ggh$ and $\gamma\gamma h$ effective coupling as functions of the modified charm Yukawa coupling is given in \autoref{app:kgkgamma}.

\section{Signal and Background}
\label{sec:signal}

\begin{figure}[htbp]
    \centering
    \begin{subfigure}{0.3\textwidth}
        \centering
        \includegraphics[width=\linewidth]{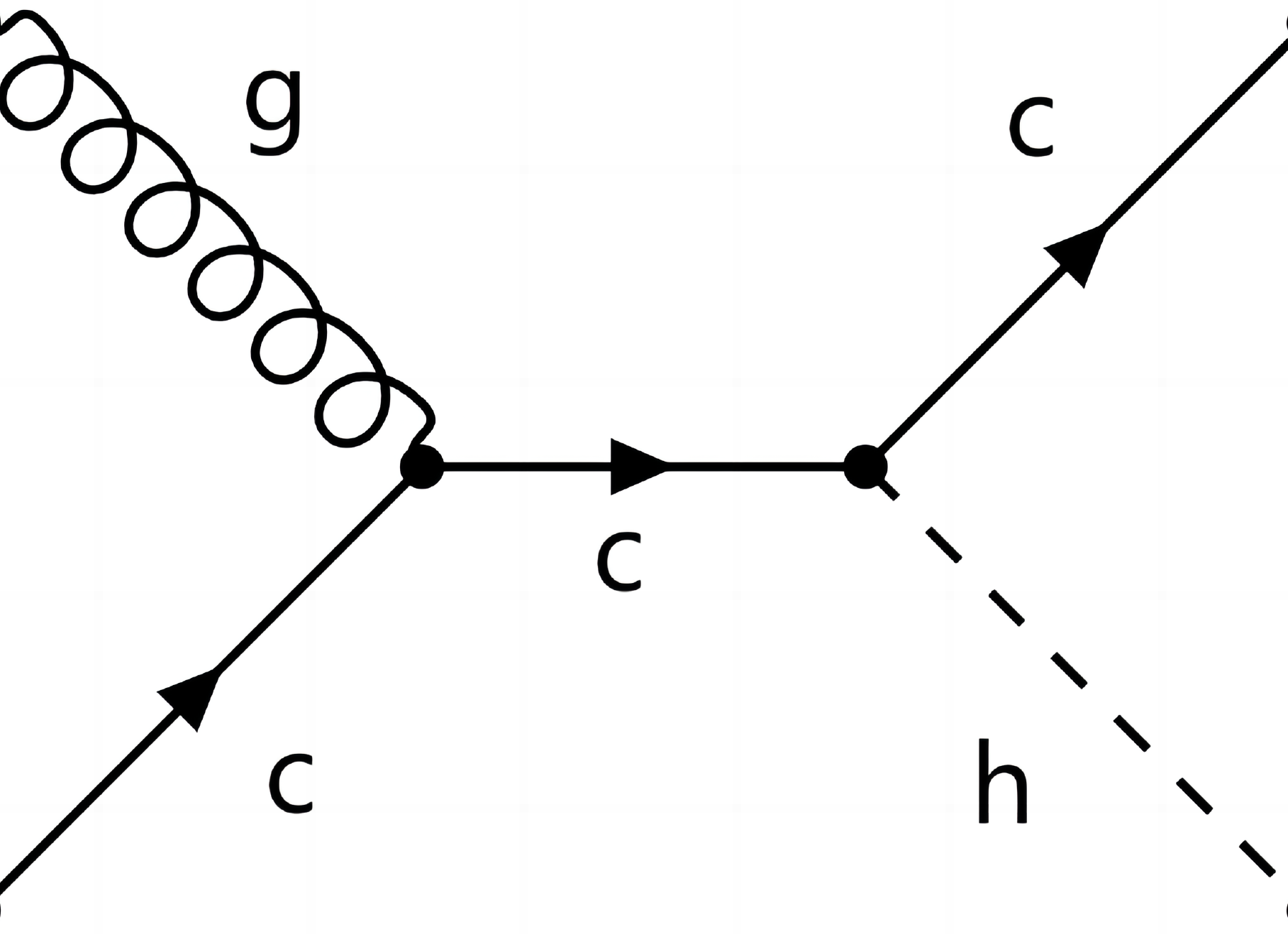}
        \caption{}
        \label{fig:feynman-diagrams1}
    \end{subfigure}
    \hfill
    \begin{subfigure}{0.2\textwidth}
        \centering
        \includegraphics[width=\linewidth]{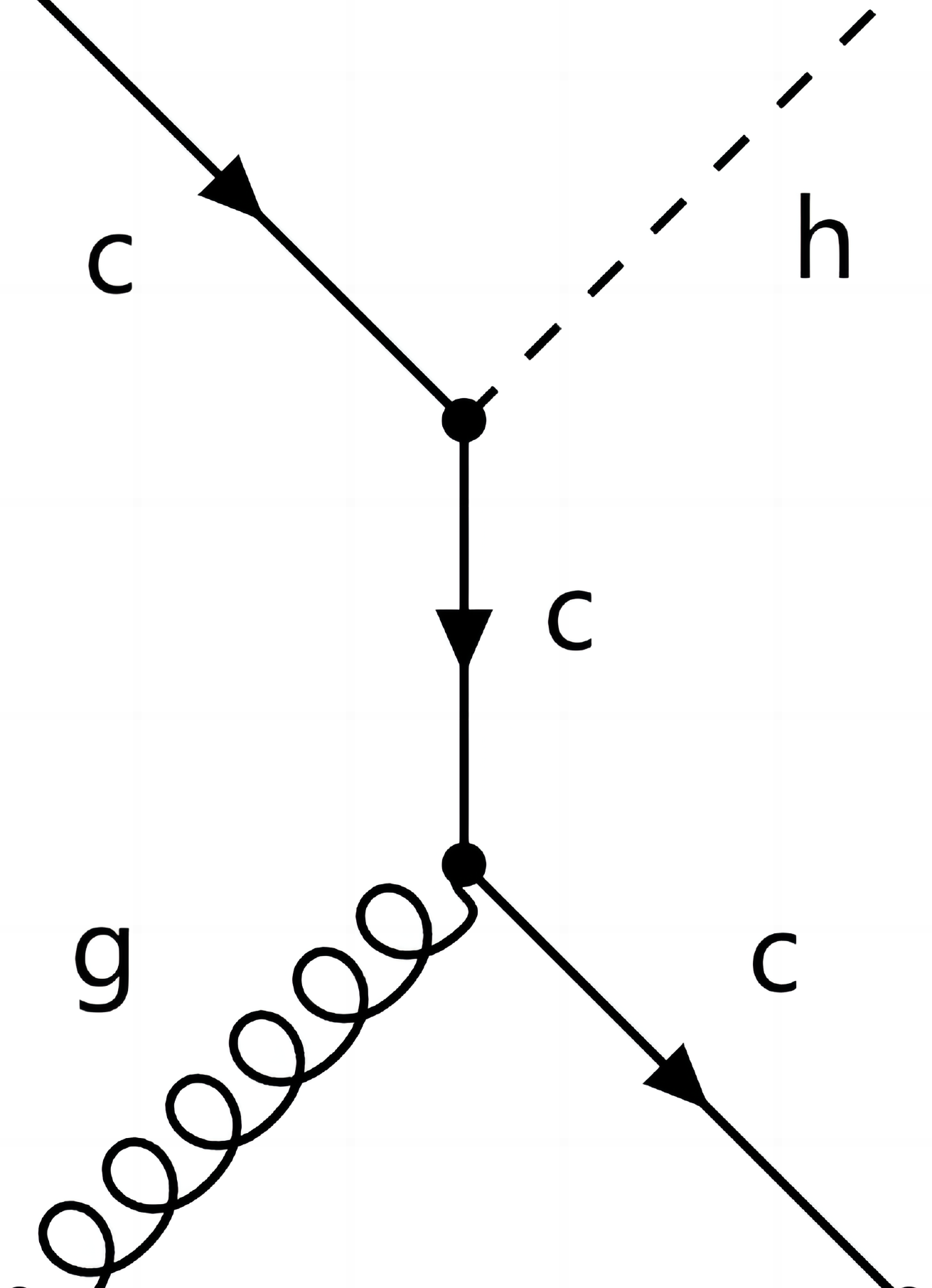}
        \caption{}
        \label{fig:feynman-diagrams2}
    \end{subfigure}
    \hfill
    \begin{subfigure}{0.2\textwidth}
        \centering
        \includegraphics[width=\linewidth]{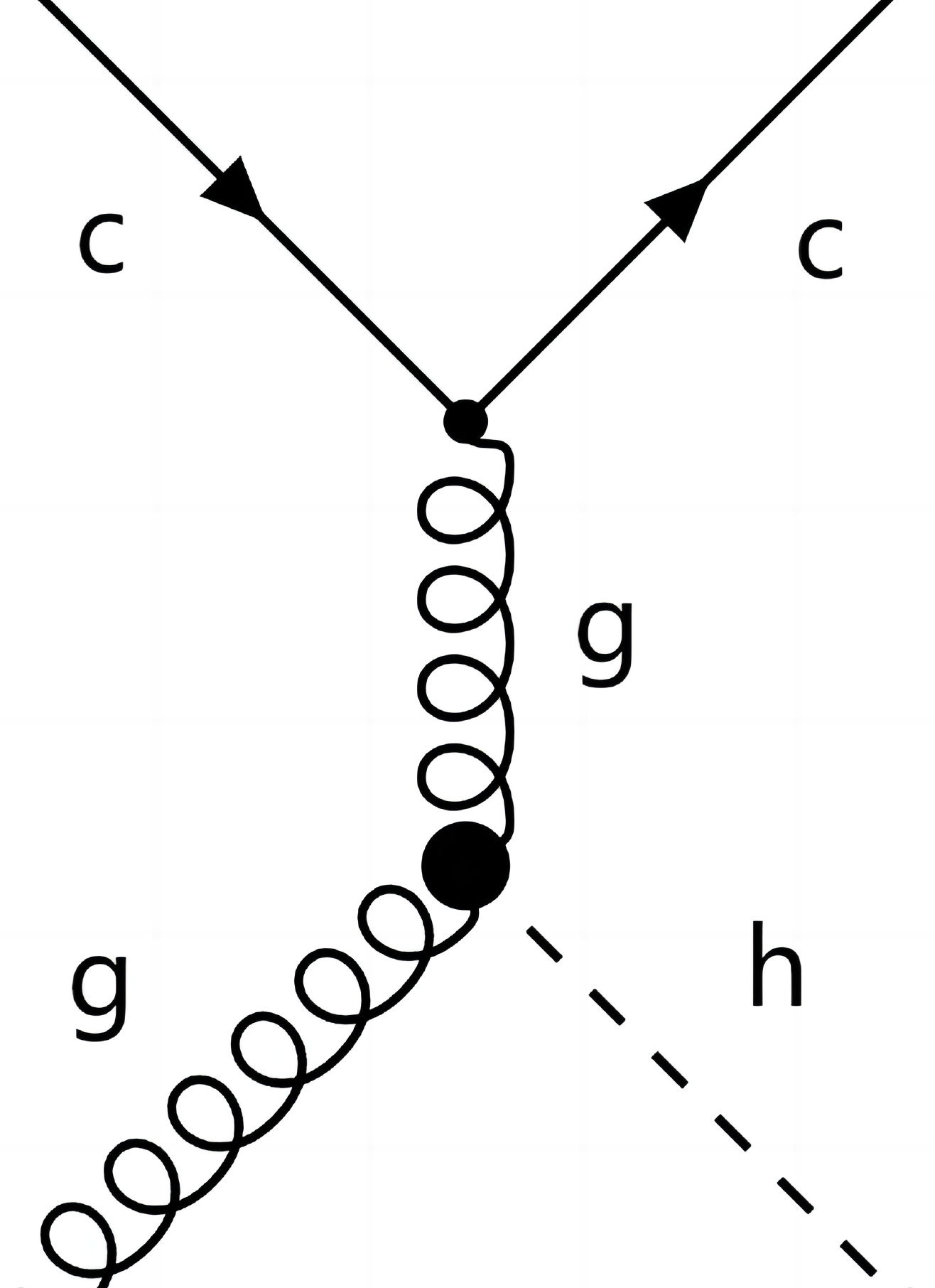}
        \caption{}
        \label{fig:feynman-diagrams3}
    \end{subfigure}
    \caption{The leading-order contribution Feynman diagrams for $pp \rightarrow hc$. Diagrams (a) and (b) depict typical diagrams involving the $cch$ Yukawa coupling, while diagram (c) illustrates contribution mediated by the $ggh$ effective coupling. The corresponding $\bar c$ diagrams are implied and not explicitly shown.}
    \label{fig:feynman-diagrams}
\end{figure}

With a four-flavor scheme Parton distribution function (PDF) and treating $ggh$ as an effective vertex $G_{\mu\nu} G^{\mu\nu}\Phi$, the leading order (LO) contribution to the signal comes mainly from two types of diagrams. The first type involves the charm Yukawa coupling, depicted in Fig.~\ref{fig:feynman-diagrams1}, \ref{fig:feynman-diagrams2}, whose amplitude is proportional to $y_c$. The second type involves the $ggh$ effective coupling, shown in Fig.~\ref{fig:feynman-diagrams3}. With a dominant top loop contribution, such diagram involves minimal dependence on light quark Yukawa, which is summarized in Appendix \ref{app:kgkgamma} and can be considered almost background-like. 
Moreover, these two types of Feynman diagrams exhibit interference effects. Within the total cross section, we designate the $|M1+M2|^2$ contribution from the first two diagrams as $cch$, which is proportional to $y_c^2$. the $|M3|^2$ term is designated as $ggh$, which is mostly independent of $y_c$. The interference term, $2{\rm Re}\left[(M1+M2)^*M3\right]$ is designated as $int$ and is proportional to $y_c$.
In the SM, the predicted contribution from $ggh$ diagram is significantly larger, by about an order of magnitude, compared to $cch$ and the $int$ term. This implies that while the signal $ch$ has a sizable total cross-section, its sensitivity  to $y_c$ is suppressed. Additionally, if the coupling contains CP phase angle, it shows in the interference part and affect the cross section as well.

As a first attempt for a realistic simulation and analysis of the $ch$ associated production signal, we chose the di-photon decay channel of the Higgs. The observed final states are $c\gamma\gamma$, where $c$ is a $c$-tagged jet which originates from a $c$ or $\bar c$ quark. \footnote{The alternative $\bar c$ quark involving processes are always implied and included, which is not to be explicitly repeated in the following. }
With this final state, the dominant irreducible background contribution is from the QCD-QED processes.
Additionally, there are fake background
such as $b\gamma \gamma ,j\gamma \gamma(j= u,d,s,g)$ where the jet is mistagged as a $c$ jet. Following the c-tagging strategy $(c\to41\%,b\to50\%,j\to3.3\%)$ from Ref.~\cite{Han:2018juw,ATLAS:2018mgv}, the cross-sections of $b\gamma \gamma$ and $j\gamma \gamma$ are less than 25$\%$ of that of $c\gamma \gamma$. Therefore, we do not include them in the full simulation of background events. 
When we define our signal to be inclusive with additional jets in the final states, contribution from the process $p p \to W  h  \to  c j \gamma \gamma$ become non-negligible. The main distinction is that this electroweak background contains two relatively hard jets with their invariant mass $m_{cj}$ peaked around the $W$ boson mass, whereas for the signal a second jet is mainly from higher order radiation which is soft. In the detailed simulation below, we include both the dominant irreducible QCD-QED background, and this large electroweak background $Wh$.

\subsection{Simulation and Kinematic cuts}

The leading-order events for both signal and background are generated using \texttt{MG5@MCNLO v3.5.3}~\cite{Alwall:2014hca}, and the parton-level events are showered and hadronized using \texttt{Pythia 8}~\cite{Sjostrand:2014zea}. Detector simulation is performed using \texttt{Delphes}~\cite{deFavereau:2013fsa}. We use \texttt{NNPDF31\_lo\_as\_0118}~\cite{Buckley:2014ana} for simulations at both the HL-LHC with a center-of-mass energy of 14 TeV and the FCC-hh with a center-of-mass energy of 100 TeV. For the input parameters, the mass of the Higgs boson are set at $M_h$ = 125 GeV, and the pole mass of the charm quark as $m_c$ = 1.3 GeV. The charm Yukawa coupling at the Higgs mass scale is set according to the running mass $\bar m_c(M_h)$ = 0.81 GeV. The renormalization and factorization scales are set at $M_h$ as in Ref.~\cite{Bizon:2021nvf}. 

In the final event selection, jets are reconstructed using the anti-kt algorithm with $\Delta R = 0.4$. One charm-tagged hadronic jet is required to satisfy $|\eta_c| < 2.5$ , $|p_{T,c}| > 20$ GeV.
We also require two photons in the final states according to detector configuration defined with corresponding HL-LHC and FCC Delphes card.
Additionally, to improve the efficiency of event generation, we require: 115 GeV$ < m_{\gamma \gamma} < 135$ GeV(HL-LHC), 118 GeV $< m_{\gamma \gamma} < 132$ GeV(FCC-hh).
We leave additional signal and background classification to BDT analysis with input observables to be defined in the next section.
We summarize the cross section and corresponding event numbers after these basic selection cuts in \autoref{tab:xsec}. Notably, the cross section contribution from interference of the $cch$ and $ggh$ diagrams is negative. 

\begin{table}
  \centering
  \begin{tabular}{|c|c|c|c|c|}
    \hline
    channel & 14 TeV $\sigma$ (fb) & 6 ab$^{-1}$ \# & 100 TeV $\sigma$ (fb) & 30 ab$^{-1} \#$  \\
    \hline
    $ c\gamma \gamma$ & 73.2 & 439,387 &  897.59  & 269,279,58  \\
    \hline
    $ Wh $ & 0.052 & 314 &  0.31 & 9,258 \\
    \hline
    $ ggh $ & 0.046 & 276 &  0.94 & 28,218 \\
    \hline
    $ cch $ & 0.0033 & 20 &  0.067  & 2,023 \\
    \hline
    $ int $ & -0.00033 & -2 &  -0.0074 & -223 \\
    \hline
  \end{tabular}
  \caption{The total cross sections and corresponding number of events for the two types of background $c\gamma\gamma$, $Wh$ and the three types of signal $ggh$, $cch$, $int$ contributions. Full detector simulation, flavor tagging strategy and basic selection cuts are applied at the HL-LHC and the FCC-hh, respectively.}
  \label{tab:xsec}
\end{table}

We would also like to add a note on correction from next leading order (NLO) contributions. The NLO introduces a new set of diagrams involving $gg , cq , cc$ and $c\bar c$ initial-states, leading to a significant enhancement to the LO cross-section. 
So far, the NLO simulation is available for the $cch$ mediated diagrams at \texttt{MG5@MCNLO}. The $ggh$ NLO simulation can be obtained using available calculation of the $h+j$@NLO process from \texttt{MCFM}~\cite{Ravindran:2002dc,Schmidt:1997wr}. For the interference term, the calculation is done in Ref.~\cite{Bizon:2021nvf}, with the code not publicly available. In our simulation, we thus stay with LO calculation at parton level using \texttt{MG5@MCNLO}, and include only the QCD real radiation shape effects with parton shower. While we differ such involved study including NLO correction to future work, we would expect a better sensitivity on the charm Yukawa coupling with improved statistics after including the full NLO correction. The methodology and overall conclusion we obtain here with LO simulation should however remain the same. 

\subsection{Observables}

To perform interpretable and full analysis on the simulated events, we need to construct a complete set of physical observables. First considering the degrees of freedom involving a three particle final-state, we have 6 observables assuming momentum conservation. $E_{cm}$ denotes the invariant mass of the $c\gamma\gamma$ final state system. $\eta_c$ and $\phi_c$ are the rapidity and the azimuthal angle of the outgoing charm jet. $\cos(\theta_{\gamma1})$ is the polar angle of the outgoing photon with the leading transverse momentum. The angle between the $cg\to ch$ interaction plane and the Higgs decay plane in the center of mass frame is denoted as $\Delta\phi$. $m_{\gamma\gamma}$ is the invariant mass of the two leading $p_T$ photons. Additionally, the momentum along the beam-direction of the $c\gamma\gamma$ system $p_{z-all}$ encodes the PDF asymmetry of the incoming partons. The transverse momentum $p_{T-all}$ and azimuthal angle $\phi_{all}$ of the $c\gamma\gamma$ system, as well as the scalar sum of transverse momenta $H_T$ of all observed final states, are defined to account for showering effects. 
To effectively reduce the $Wh$ background contribution, we also include an observable $m_{cj}$, the invariant mass of the charm jet and the hardest $p_T$ jet identified without a flavor tag. In all, we choose an over-complete set of observables for a final state of $c\gamma\gamma + X$, inclusive with additional jets. By calculating the correlation matrix among the observations, we find that most correlations are negligible, with only $E_{cm}, H_T $, $p_{T-all}$ and $m_{cj}$ showing high levels of correlation at the level of about 30\%. This is reasonable to expect, as they all relate to the overall energy scale of the collision event, yet provide marginally additional information for distinguishing between signal and background processes.

\section{\texorpdfstring{$pp \to ch$}{pp -> ch} at Future Colliders}
\label{sec:analysis}
As mentioned earlier, we categorize contribution to the total $c\gamma\gamma$ cross section into three types of $ch$ signals and two types of non-Higgs backgrounds, an overall five categories. The Physics goal is to evaluate the sensitivity to both the magnitude and a possible CP-phase of the charm Yukawa, we aim to distinguish the $cch$ and $int$ contribution from all. Thus we must address the challenge of optimizing a multi-class discrimination based on multi-observable inputs. We use the BDT algorithm implemented in XGBoost~\cite{Chen:2016btl} in our analysis. We include the over-complete set of observables as inputs to the BDT training, to ensure all physics information known at the simulation level are exploited. 
In recent years, interpretable machine learning has gained significant popularity in high-energy physics and offered additional insights and understanding in collider signal analysis. Therefore, we use shapley values~\cite{RM-670-PR} in the analysis to dissect the optimization results quantitatively.  
The efficiency of this approach are demonstrated in Ref.~\cite{Grojean:2020ech}. After training the BDT model, we calculate the shapley values based on the BDT network, and then interpret the averaged absolute shapley values for each observable ($\bar {\left| S_v \right| }$) as their importance in making the optimized categorization. The larger the $\bar {\left| S_v \right| }$ value of a certain observable, the greater its contribution to distinguish different signal and background channel contributions.
We show the main analysis results at the two collider settings in the following.

\subsection{HL-LHC}

\begin{figure}[htbp]
    \centering
    \begin{subfigure}{0.3\textwidth} % 调整需要的宽度
        \centering
        \includegraphics[width=\linewidth]{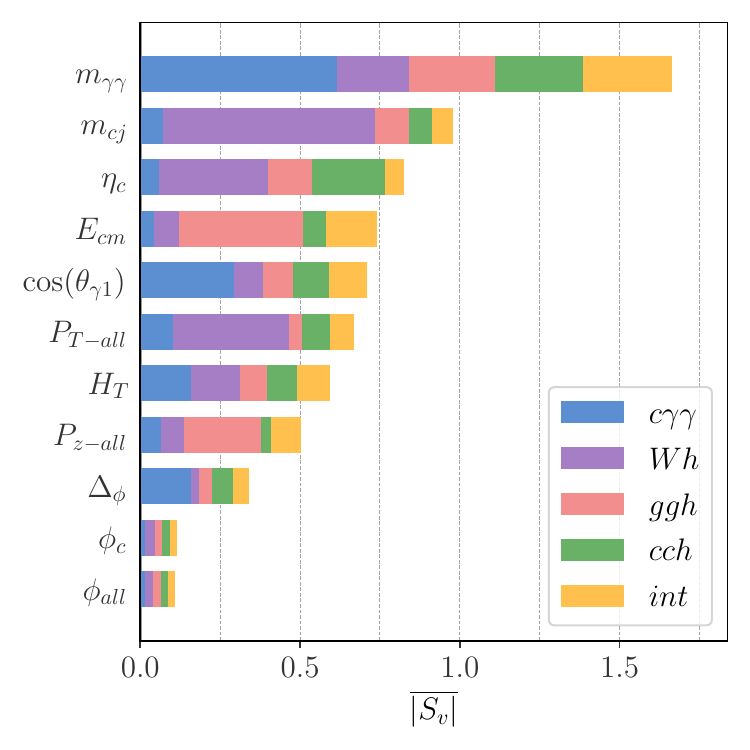}
    \end{subfigure}
    \hfill
    \begin{subfigure}{0.3\textwidth} % 调整需要的宽度
        \centering
        \includegraphics[width=\linewidth]{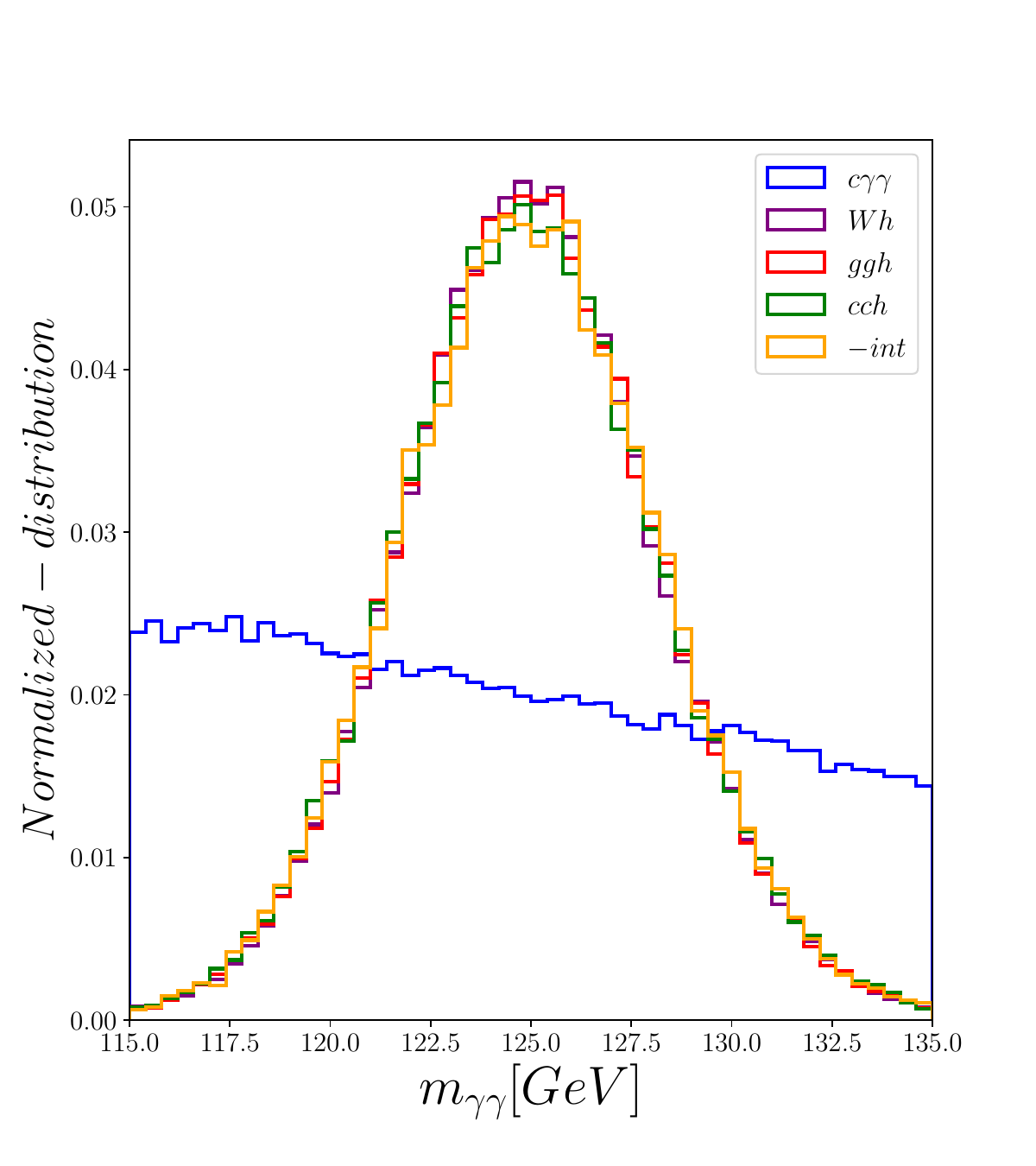}
    \end{subfigure}
    \hfill
    \begin{subfigure}{0.3\textwidth} % 调整需要的宽度
        \centering
        \includegraphics[width=\linewidth]{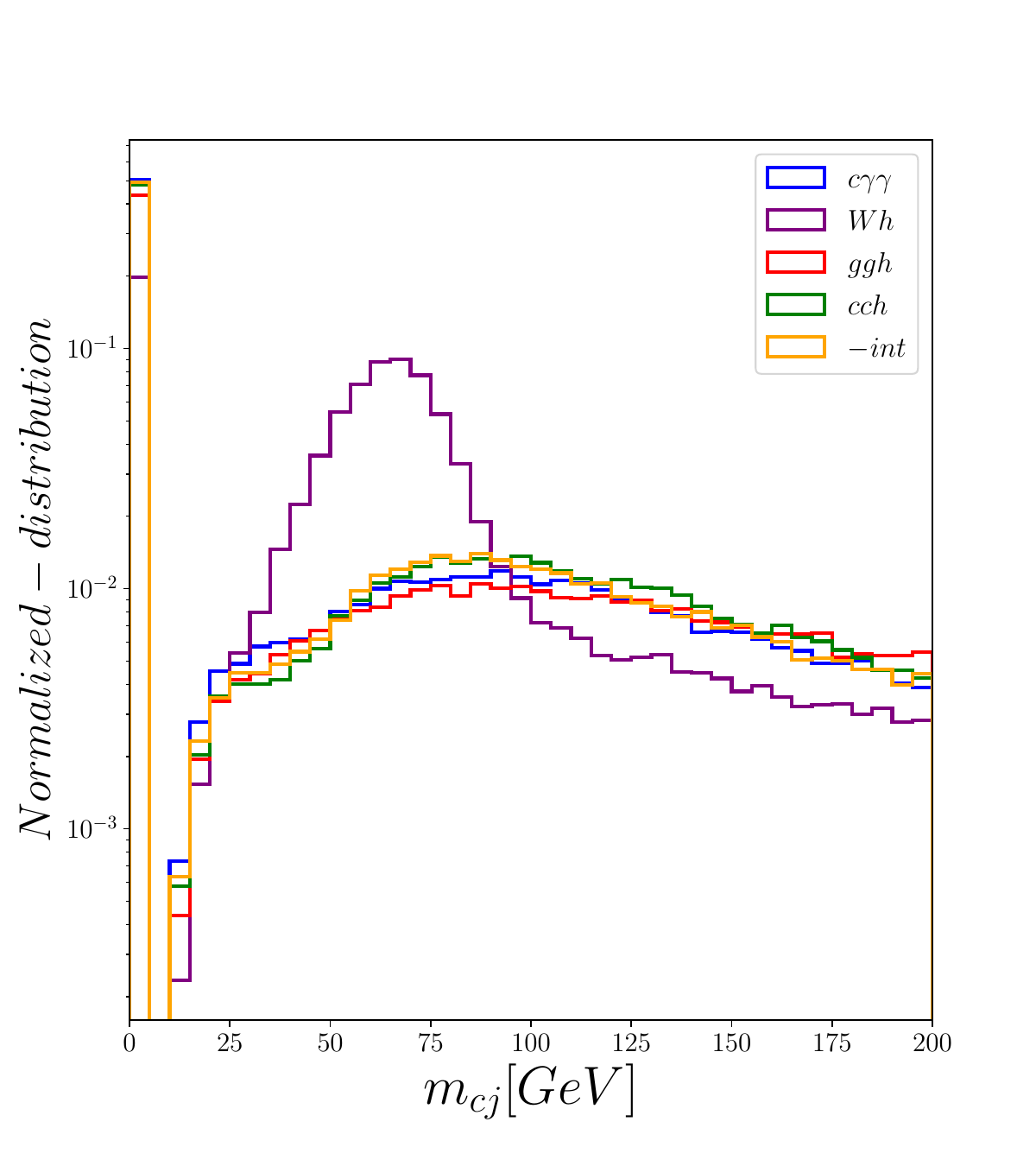}
    \end{subfigure}

    \vspace{0.1cm} % 调整上下间距
    
    \begin{subfigure}{0.3\textwidth} % 调整需要的宽度
        \centering
        \includegraphics[width=\linewidth]{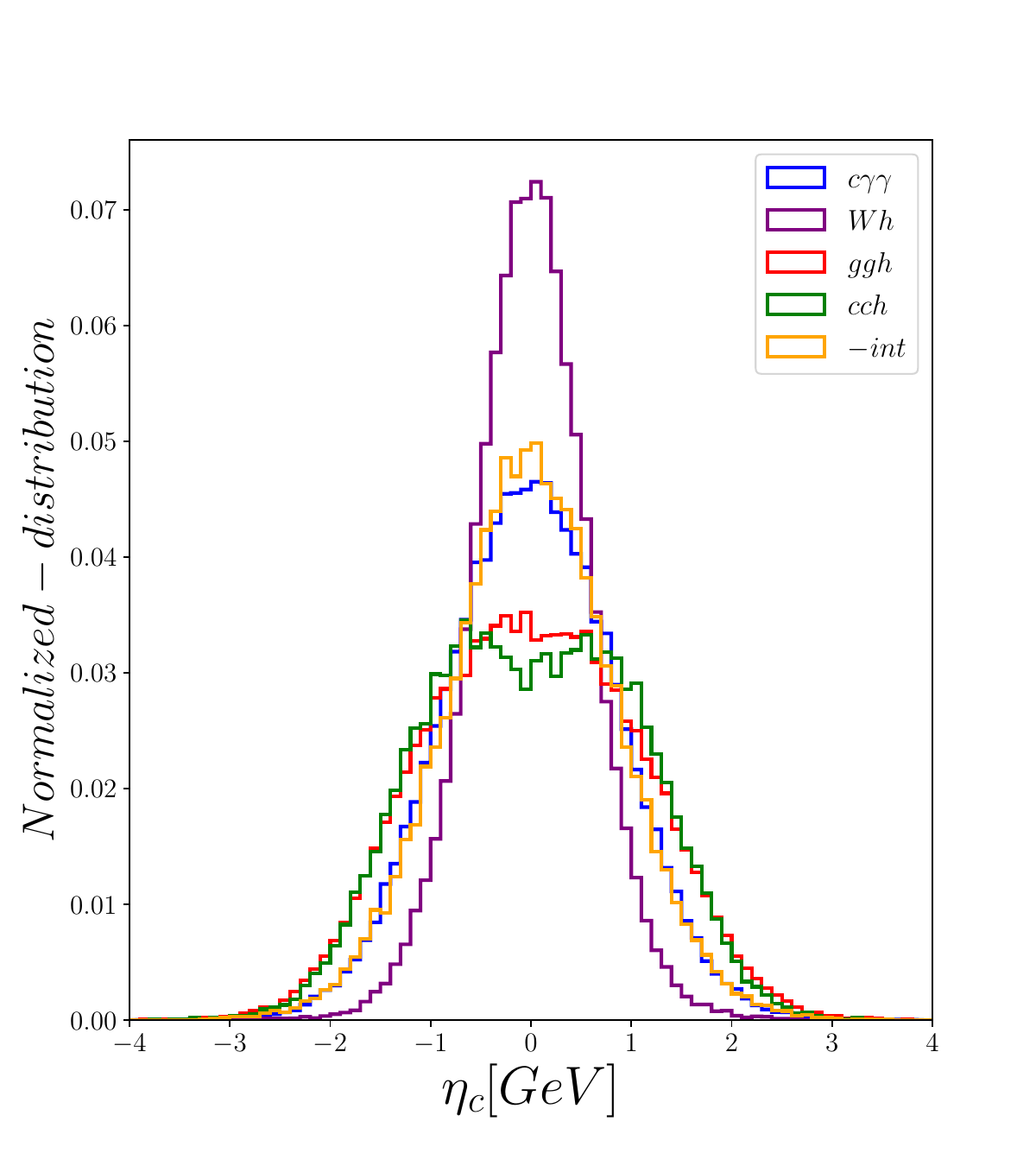}
    \end{subfigure}
    \hfill
    \begin{subfigure}{0.3\textwidth} % 调整需要的宽度
        \centering
        \includegraphics[width=\linewidth]{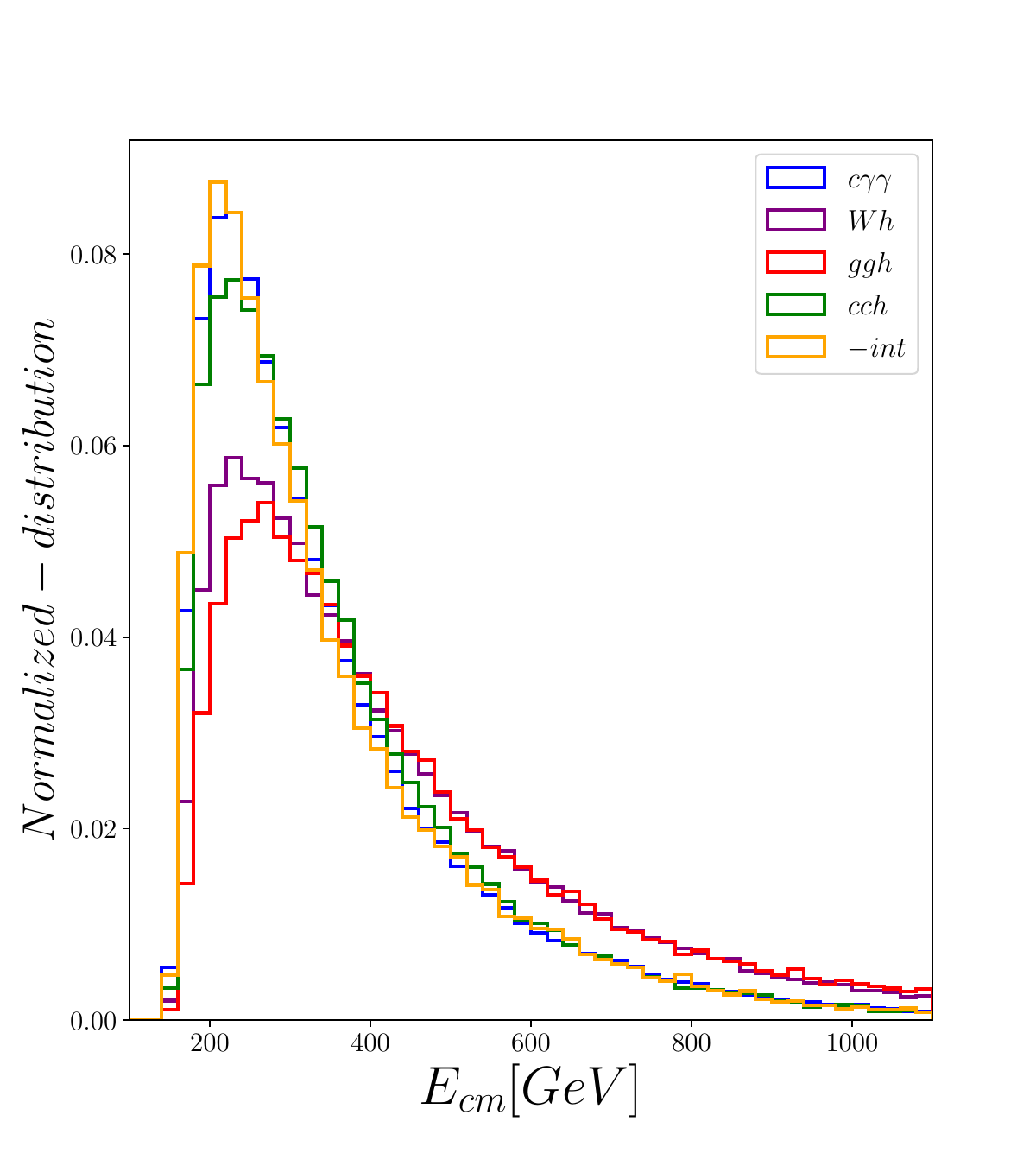}
    \end{subfigure}
    \hfill
    \begin{subfigure}{0.3\textwidth} % 调整需要的宽度
        \centering
        \includegraphics[width=\linewidth]{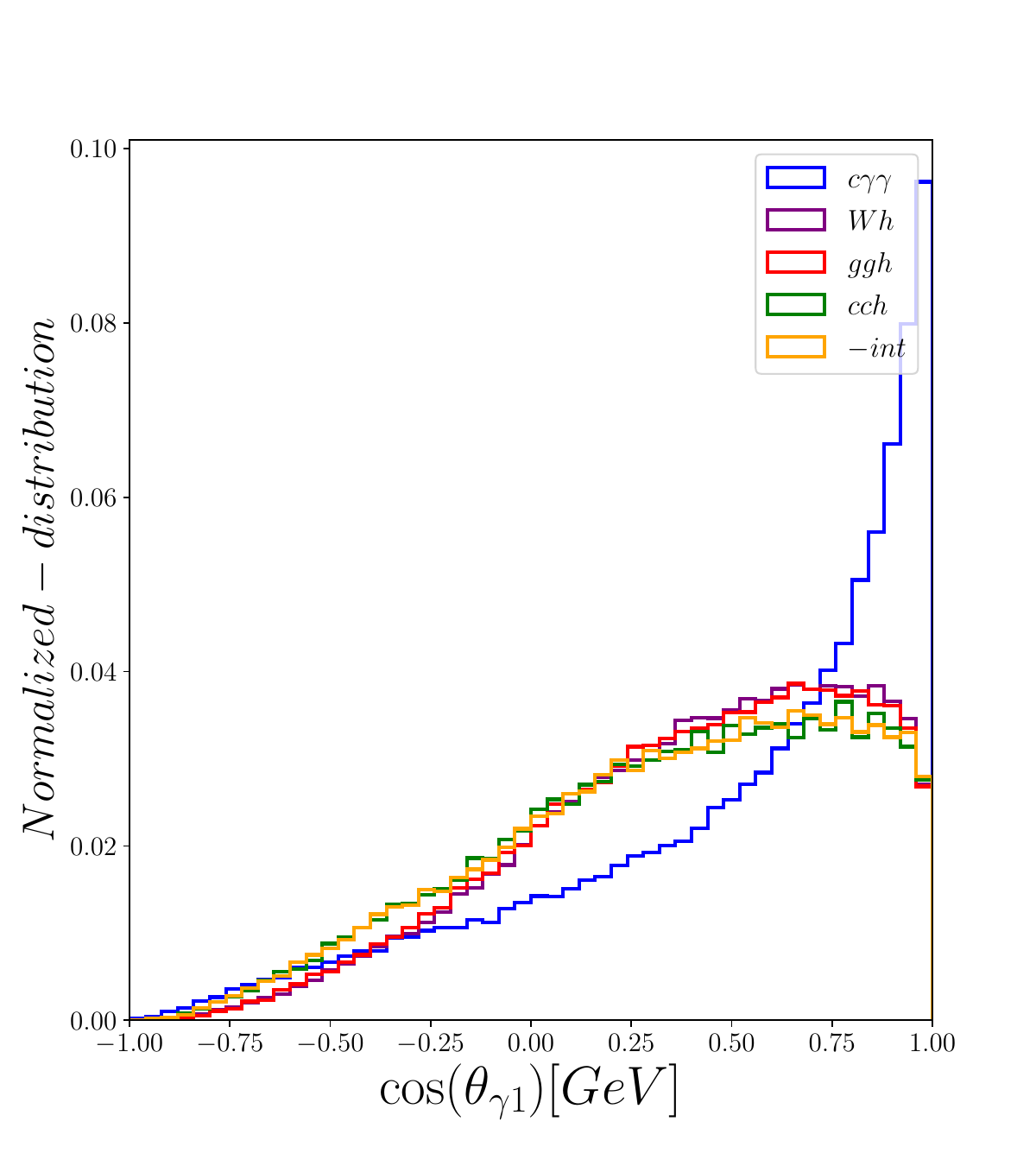}
    \end{subfigure}
    \caption{Based on simulated events at the HL-LHC and the BDT categorization analysis, shapley importance ranking and the differential distributions of the top most important kinematic variables for each channel are shown.}
    \label{fig:hllhc-observables}
\end{figure}

The importance ranking of the observables is shown in the first panel of Fig.~\ref{fig:hllhc-observables}. In order to understand the importance ranking obtained from the BDT analysis, we also include the differential cross section distributions of the top most important observables in the rest of Fig.~\ref{fig:hllhc-observables}. In the importance ranking plot, we show the mean of absolute shapley value $\bar {\left| S_v \right| }$ representing different channels with different colors. The length quantify the contributing importance of that observable from distinguishing that specific channel.

Taking $m_{\gamma \gamma}$ as an example, since the di-photon from the three signal channels as well as the $Wh$ background are all from the Higgs decay, their invariant mass distribution has the same shape peaking at the Higgs mass, while the background $c\gamma\gamma$ exhibits a relatively flat distribution. This observable naturally plays the most important role in distinguishing these two types of contributions. In the $\bar {\left| S_v \right| }$ shapley importance ranking, the score of $m_{\gamma \gamma}$ from $c \gamma \gamma$ is the highest, while the scores for the other three channels are similar. The $m_{cj}$ distribution from $Wh$ channel has an expected wide peak below the $W$ boson mass and distinguishes the channel from the others effectively. As expected, $m_{cj}$ importance ranking has dominant contribution from the $Wh$ channel. For the observable $\eta_c$, we observe that it has distinctive shape for the $cch$ and $Wh$ channels from the others. The $cch$ contribution exhibits a relative plateau and small dent at small absolute values, while the others all peak at the center. Additionally, the sharpness of the peak differentiates $Wh$ from the rest of the channels. It thus offers high distinction power for $cch$ and $Wh$ in the importance ranking. Similar features can be observed and understood in the importance ranking for the other observables as well.

After understanding the importance of observables in differentiating different contributing channels, we proceed to use the trained BDT model to predict and classify the simulated data. Since this analysis involves five channels, we present the predictions of the BDT using a 5×5 confusion matrix. The element at position $(i, j)$ in the matrix represents the number of occurrences of the ``real (simulated)" $i^{th}$ event type being predicted within $j^{th}$ event category. The confusion matrix for HL-LHC is presented in Table \ref{tab:bdtclass-lhc}. We choose a commonly used significance formula $(\sigma = N_S/\sqrt{N_S + N_B })$ to represent the corresponding ``signal" strength within each predicted category. In the confusion matrix, we define it as:
\begin{equation}
     \sigma_i = \frac{|N_{ii}|}{\sqrt{\sum_j N_{ij}}}  
\end{equation}

At HL-LHC, the predicted number of events for the $cch$ and $int$ channel is small given limited cross section. The significance obtained for the $cch$ channel is 0.03$\sigma$. For the $int$ channel we lose sensitivity for the phase angle. Based on this, we can still provide a constraint on a real modification of charm Yukawa $\kappa_c$, which is calculated in the next section.

\begin{table}
  \centering
  \begin{tabular}{c|lllll|l}
    \multicolumn{6}{c}{\textbf{\large Predicted no. of events at HL-LHC}} \\
    \hline
    Channel &  $c\gamma \gamma$ &  $wh$  &  $ggh$  & $cch$  &  $int$ & total\\
    \hline
    $c\gamma \gamma$ &  389,531 & 22,373 &  10,194 &  9,582 &  7,707 & 439,387 \\
    \hline
   $wh$    &    91 &    193 &   13 &      7 &   10  &   314 \\
    \hline
     $ggh$    &   131 &     31 &    77 &     25   &  12  & 276 \\
    \hline
    $cch$    &  12 &  2 & 2  &  3 &  1 & 20 \\
    \hline
    $int$    &  -2 &  0 & 0  &  0 &  0 & -2 \\
    \hline
    $\sigma_j$ &   623.93  &     1.28 &   0.76   &      0.031 &  0      \\
    \hline
  \end{tabular}
  \caption{Trained BDT classification (confusion matrix) of the five channel contributions at HL-LHC with an integrated luminosity of 6 ab$^{-1} $(ATLAS+CMS). The right-most column provides the total number of events expected from each true channel in the SM. The last row shows the expected significance for the corresponding channel.}
  \label{tab:bdtclass-lhc}
\end{table}

\subsection{FCC-hh}

\begin{figure}[htbp]
    \centering
    \begin{subfigure}{0.3\textwidth} % 调整需要的宽度
        \centering
        \includegraphics[width=\linewidth]{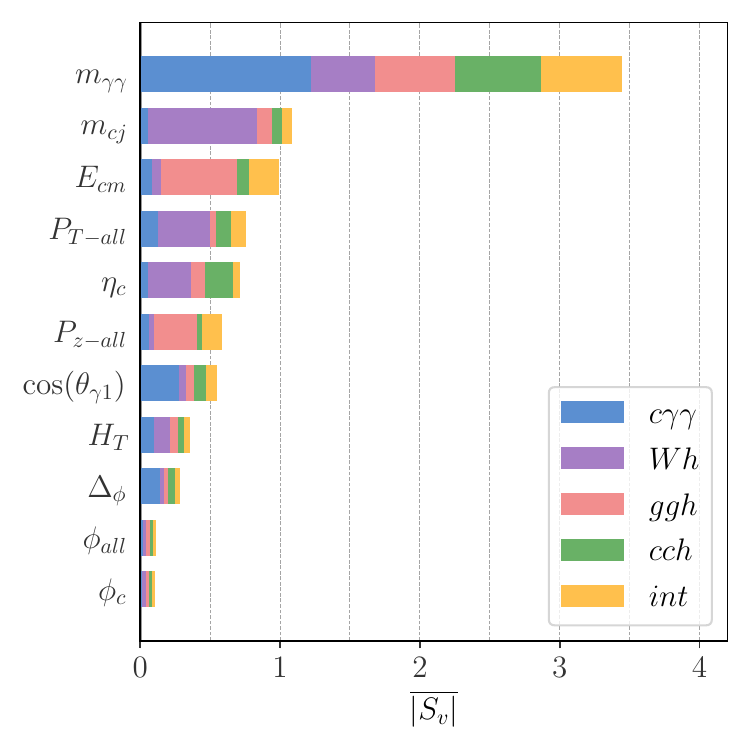}
       
    \end{subfigure}
    \hfill
    \begin{subfigure}{0.3\textwidth} % 调整需要的宽度
        \centering
        \includegraphics[width=\linewidth]{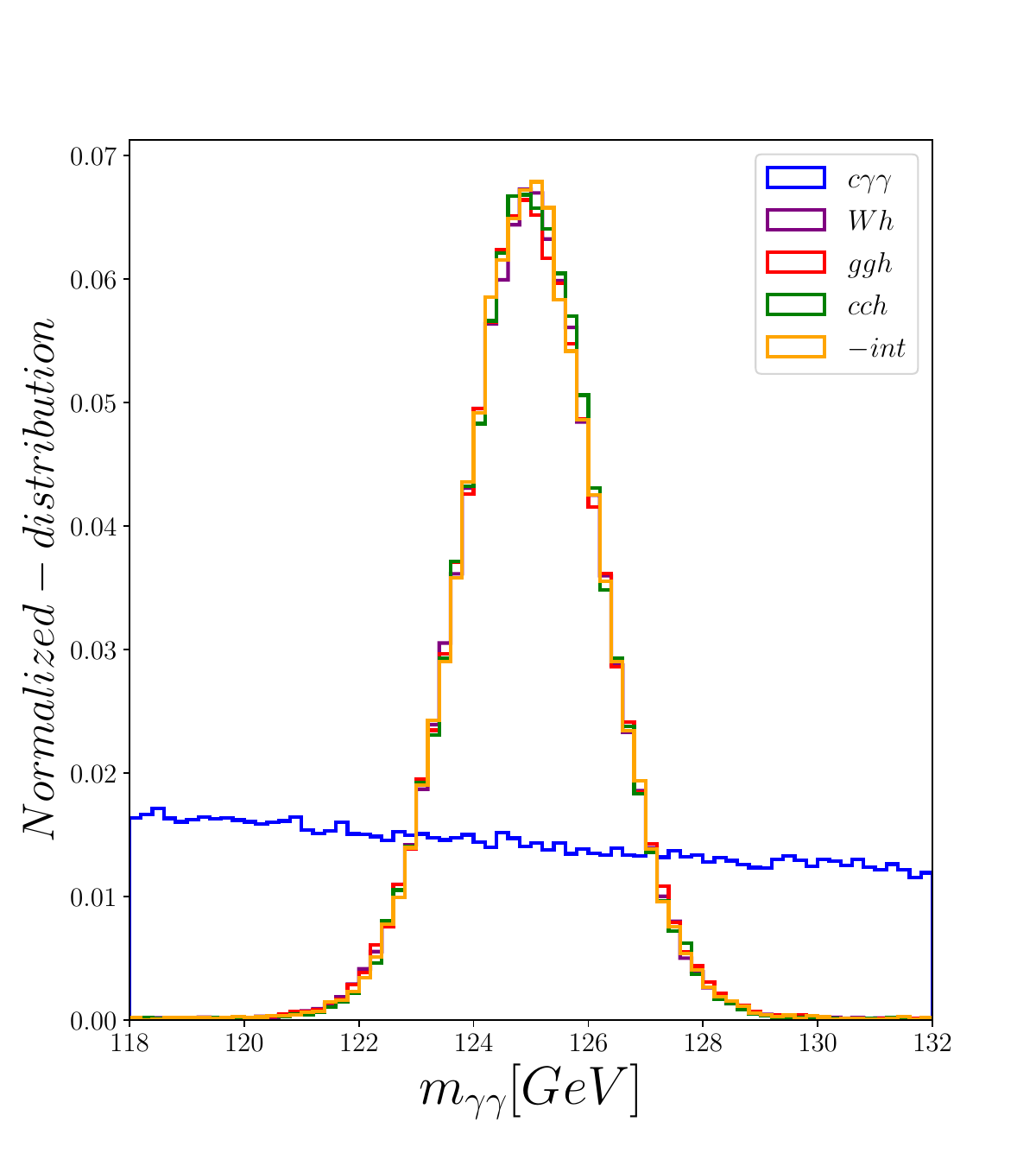}
       
    \end{subfigure}
    \hfill
    \begin{subfigure}{0.3\textwidth} % 调整需要的宽度
        \centering
        \includegraphics[width=\linewidth]{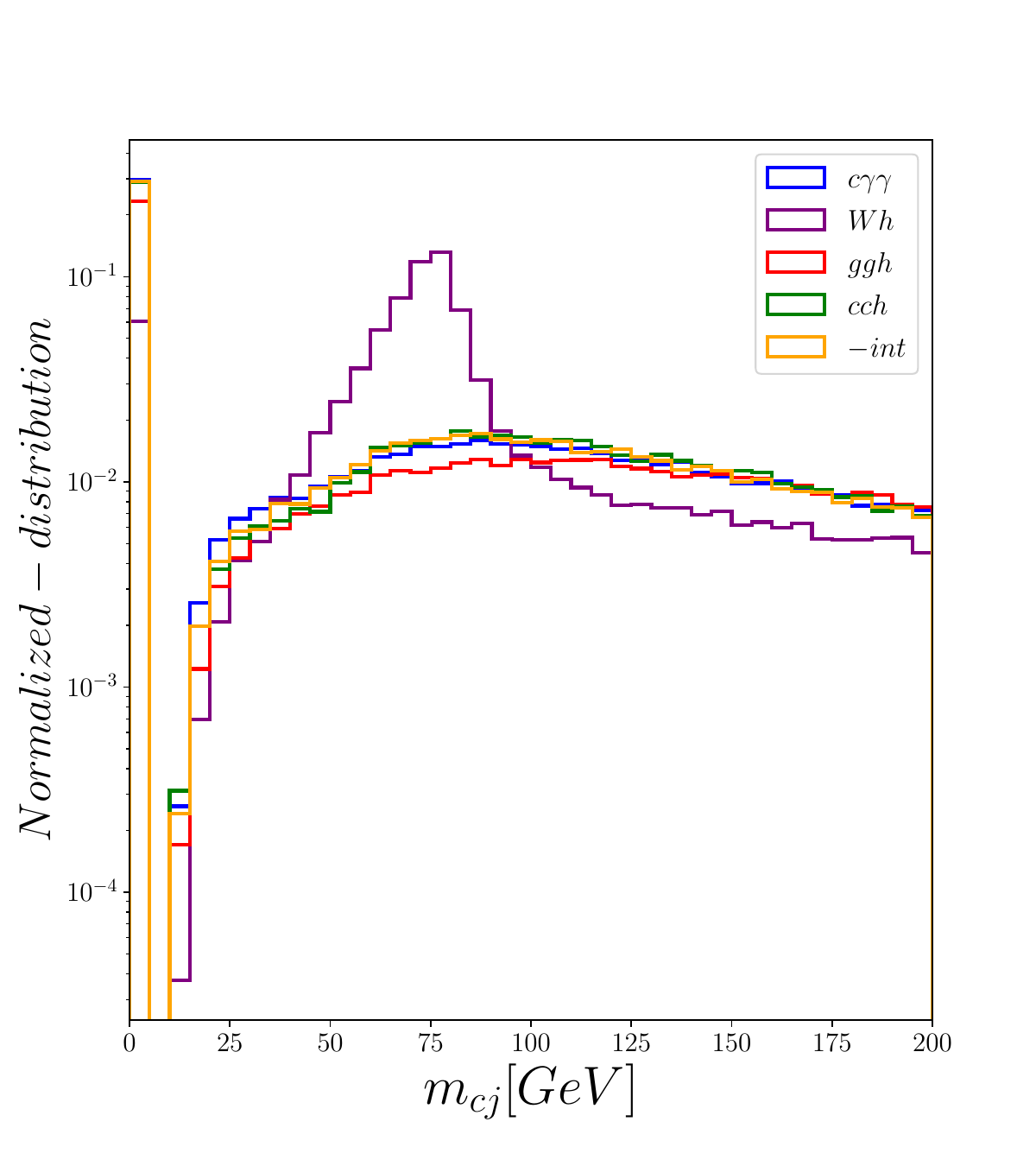}
       
    \end{subfigure}

    \vspace{0.1cm} % 调整上下间距
    
    \begin{subfigure}{0.3\textwidth} % 调整需要的宽度
        \centering
        \includegraphics[width=\linewidth]{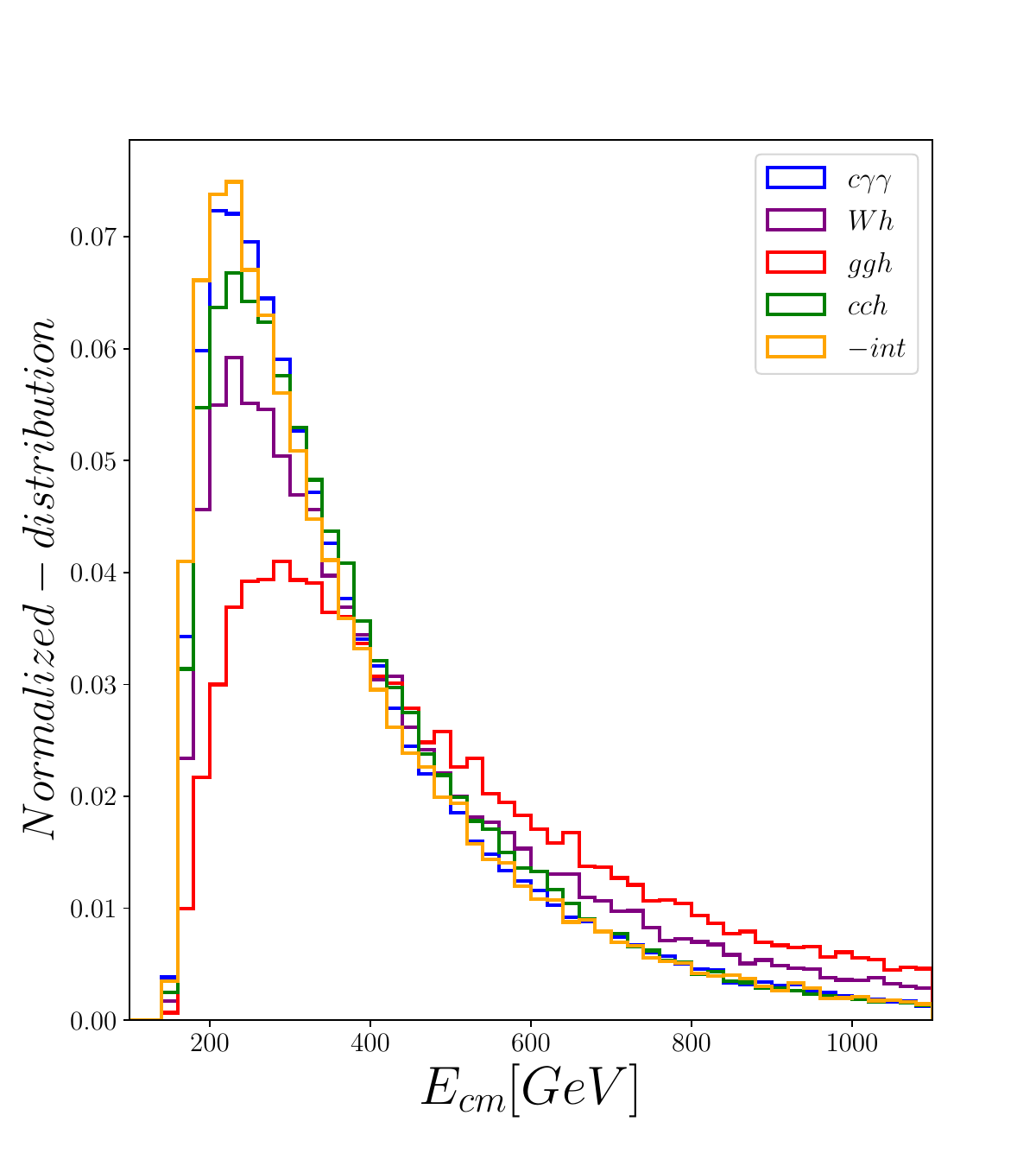}
        
    \end{subfigure}
    \hfill
    \begin{subfigure}{0.3\textwidth} % 调整需要的宽度
        \centering
        \includegraphics[width=\linewidth]{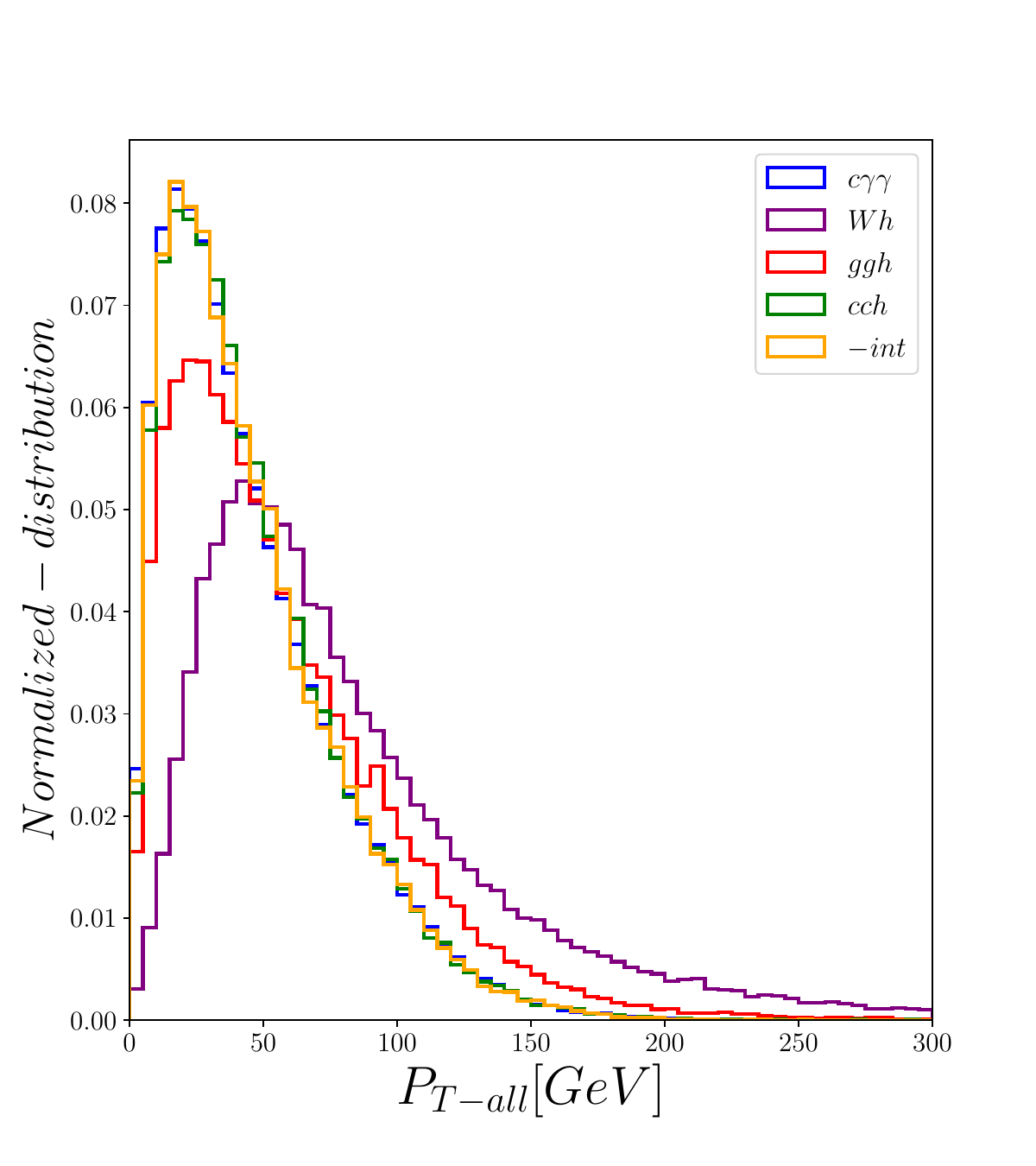}
       
    \end{subfigure}
    \hfill
    \begin{subfigure}{0.3\textwidth} % 调整需要的宽度
        \centering
        \includegraphics[width=\linewidth]{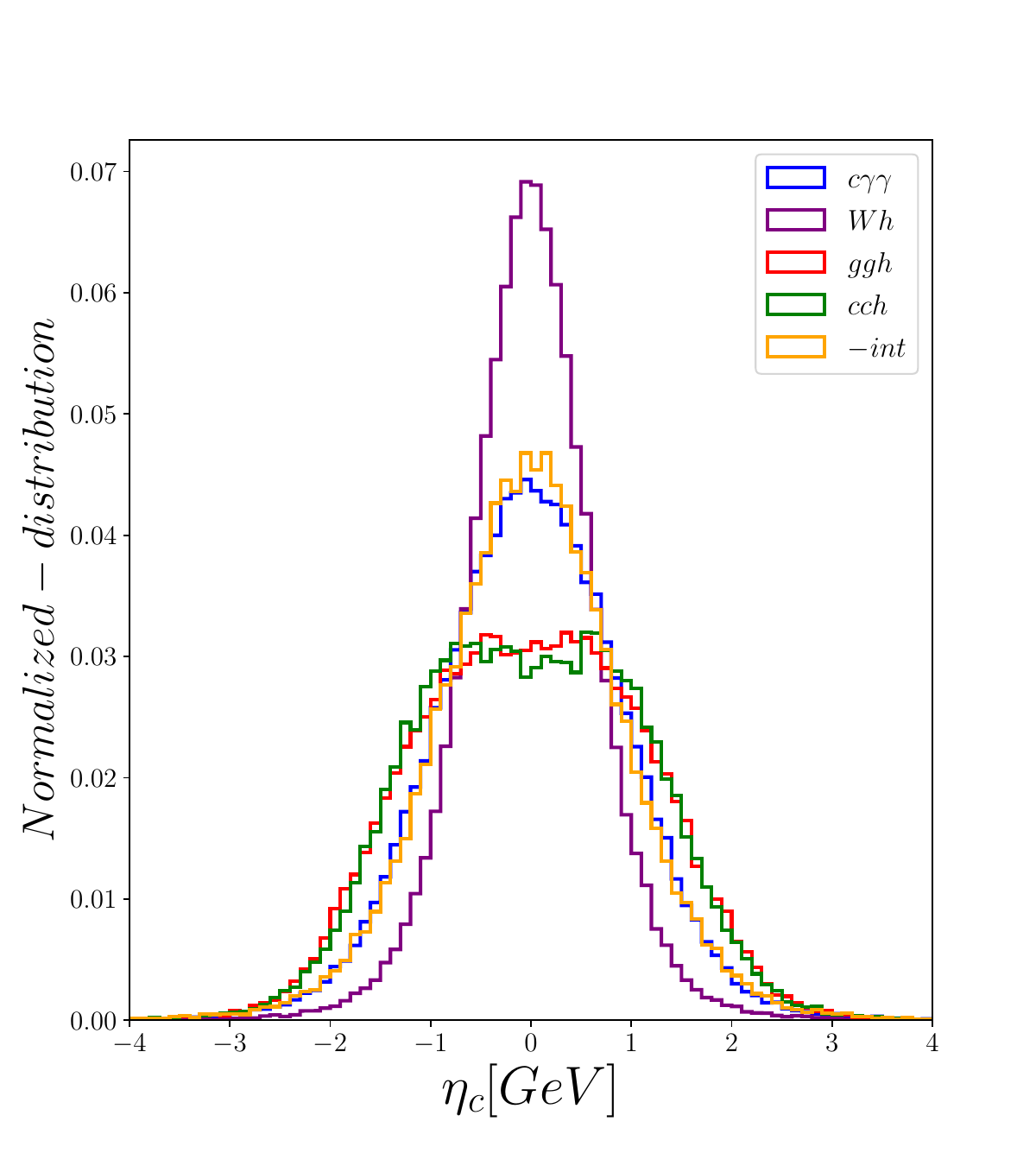}
       
    \end{subfigure}
    \caption{Based on simulated events at the FCC-hh and the BDT categorization analysis, shapley importance ranking and the differential distributions of the top most important kinematic variables for each channel are shown.}
    \label{fig:fcc-observables}
\end{figure}

As in the case for the HL-LHC, we conduct the same event simulations and analysis scheme for the future collider scenario FCC-hh, considering 100 TeV $pp$ collision with $30$ ab$^{-1}$ integrated luminosity. At the FCC-hh, the $cch$ and $int$ channel have much enhanced statistics, not only due to the increase in accumulated luminosity but also because of the increase in the cross-section. From Table~\ref{tab:xsec}, it can be observed that the cross-sections of $c\gamma \gamma ,cch$ have increased by approximately 18, $ggh$ by approximately 20, and $int$ by approximately 11 times compared to the HL-LHC.
After training the BDT model, we generated the $\bar {\left| S_v \right| }$ importance ranking and event distribution plots for the five most important observables, which are shown in Fig.~\ref{fig:fcc-observables}. The basic features remain similar to the HL-LHC where the top two most important observable remain $m_{\gamma\gamma}$ and $m_{cj}$ in helping to distinguish the dominant background process $c\gamma\gamma$ and the electroweak background $Wh$. Notably at the FCC-hh, better prospected detector resolution reconstructs a narrower Higgs mass peak from the di-photon invariant mass, making $m_{\gamma \gamma}$ much more effective in distinguishing between signal and background and much larger importance ranking score. 
The resulting confusion matrix from BDT optimized classification and the observation significance achieved for each channel are summarized in Table~\ref{tab:bdtclass-fcc}. Compared to the results in HL-LHC, the number of events in all categories increased significantly, especially with the events in the $int$ category beginning to be discernible. The significance of the $cch$ channel reaches 0.5$\sigma$, and the $int$ channel reaches 0.06$\sigma$. Although their significance levels do not reach 1$\sigma$, we derive constraints on the modified space of charm Yukawa coupling.

\begin{table}
  \centering
  \begin{tabular}{c|lllll|l}
    \multicolumn{6}{c}{\textbf{\large Predicted no. of events at FCC-hh}} \\
    \hline
    Channel &  $c\gamma \gamma$ &  $wh$  &  $ggh$  & $cch$  &  $int$ & total\\
    \hline
    $c\gamma \gamma$ & 22,748,500 & 1,056,348 & 781,197 & 1,005,789 & 1,336,122 & 26,927,958 \\
    \hline
   $wh$    &    1,282 &    6,061 &   526 &     482 &   871  &   9258 \\
    \hline
     $ggh$    &   5,994 &     2,949 &    11,478 &     4,338   &  3,459  & 28,218 \\
    \hline
    $cch$    &  529 &  221 & 311  &  516 &  446 & 2,023 \\
    \hline
    $int$    &  -64 &  -30 & -23  &  -42 &  -64 & -223 \\
    \hline
    $\sigma_j$ &   4,768  &   5.87 &   12.88   &      0.51 &  0.055      \\
    \hline
  \end{tabular}
  \caption{Trained BDT classification (confusion matrix) of the five channel contributions at FCC-hh with a luminosity of 30 ab$^{-1} $. The right-most column provides the total number of events expected from each true channel in the SM.}
  \label{tab:bdtclass-fcc}
\end{table}

\section{Constraints on the Yukawa Coupling}
\label{sec:results}
Given the analysis above, we utilize the event counts of the two signal categories from the confusion matrix obtained to constrain the Charm-Yukawa coupling. To investigate possible CP term in the Yukawa sector, we use $\kappa_c$ to describe the CP-even part and $\tilde{\kappa_c}$ the CP-odd part. Alternatively, we can use $|\kappa_c|$ and the phase angle $\alpha$ to describe the complex Yukawa coupling. 
Note that the sensitivity we get on the CP-phase depends on sensitivity achievable to the interference term between $cch$ and $ggh$ with the latter assumed to be SM. 
%Given the low event rate at the HL-LHC, we can reach meaningful bounds on possible CP-odd term in FCC-hh.

In addition to our probe of CP phase through the interference term, direct probe of the CP structure in the fermion-Higgs coupling by constructing CP sensitive observable is done in Ref.~\cite{Cassidy:2023lwd,Barman:2022pip,Mb:2022rxu,Boudjema:2015nda,Goncalves:2018agy,Biswas:2012bd,Biswas:2013xva,Ellis:2013yxa,Chang:2014rfa,CMS-PAS-HIG-14-001,CMS-PAS-HIG-16-019,ATLAS:2014ayi}. Such observable construction are so far limited to probe the top and tau Yukawa couplings where the polarization information could be retrieved through their decay kinematics. 
With the accumulation of LHC data and the expectation of increased event rate, the CP-structure information for bottom and charm Yukawa couplings may be probed with the help of jet substructure observable. This however is beyond the scope of current study and we leave for future exploration.

\subsection{Constraints on a real charm Yukawa}

\begin{figure}[htbp]
    \centering
    \begin{subfigure}{0.49\textwidth} % 调整子图宽度
        \centering
        \includegraphics[width=\linewidth]{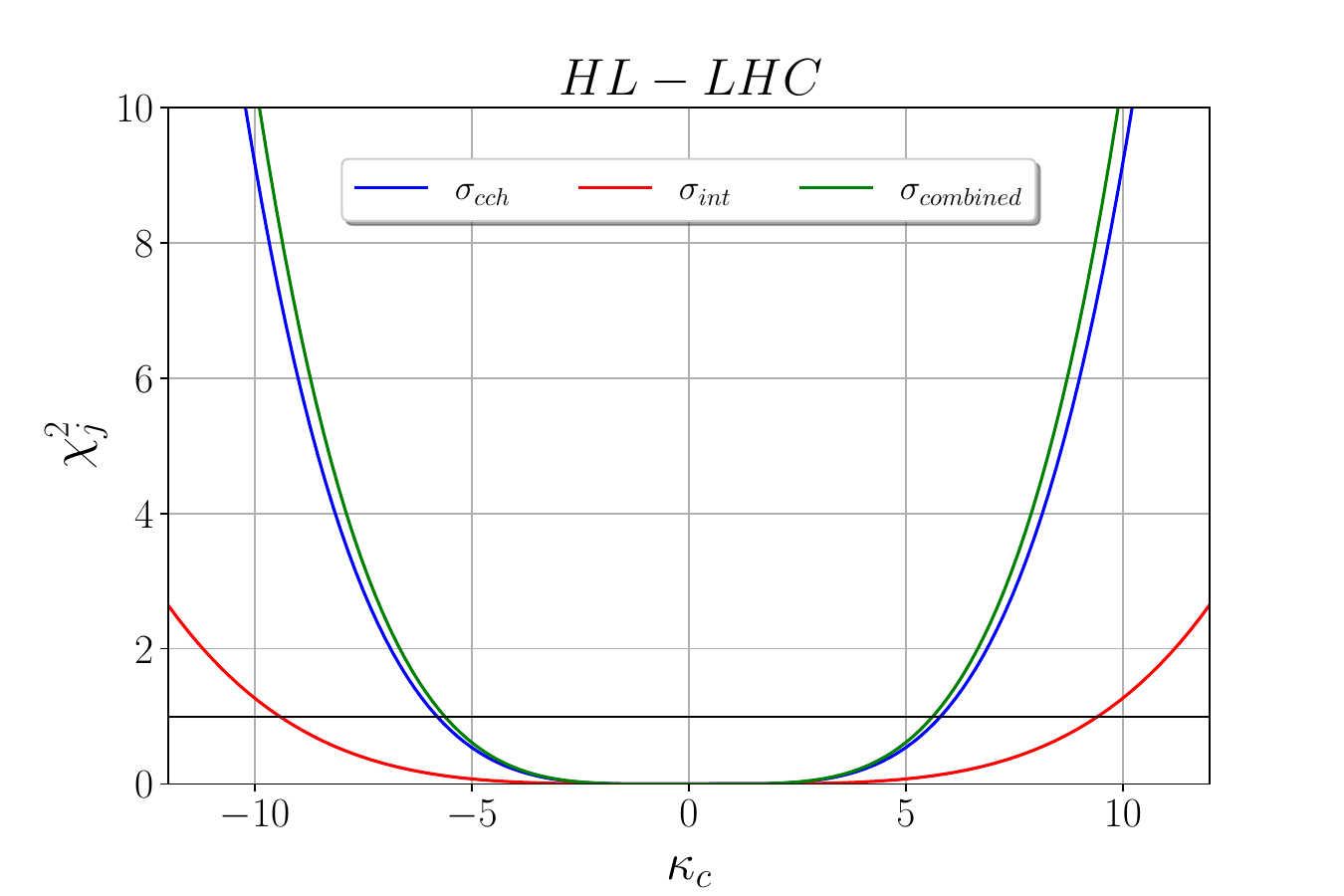}
        \caption{}
        
    \end{subfigure}
    \hfill
    \begin{subfigure}{0.49\textwidth} % 调整子图宽度
        \centering
        \includegraphics[width=\linewidth]{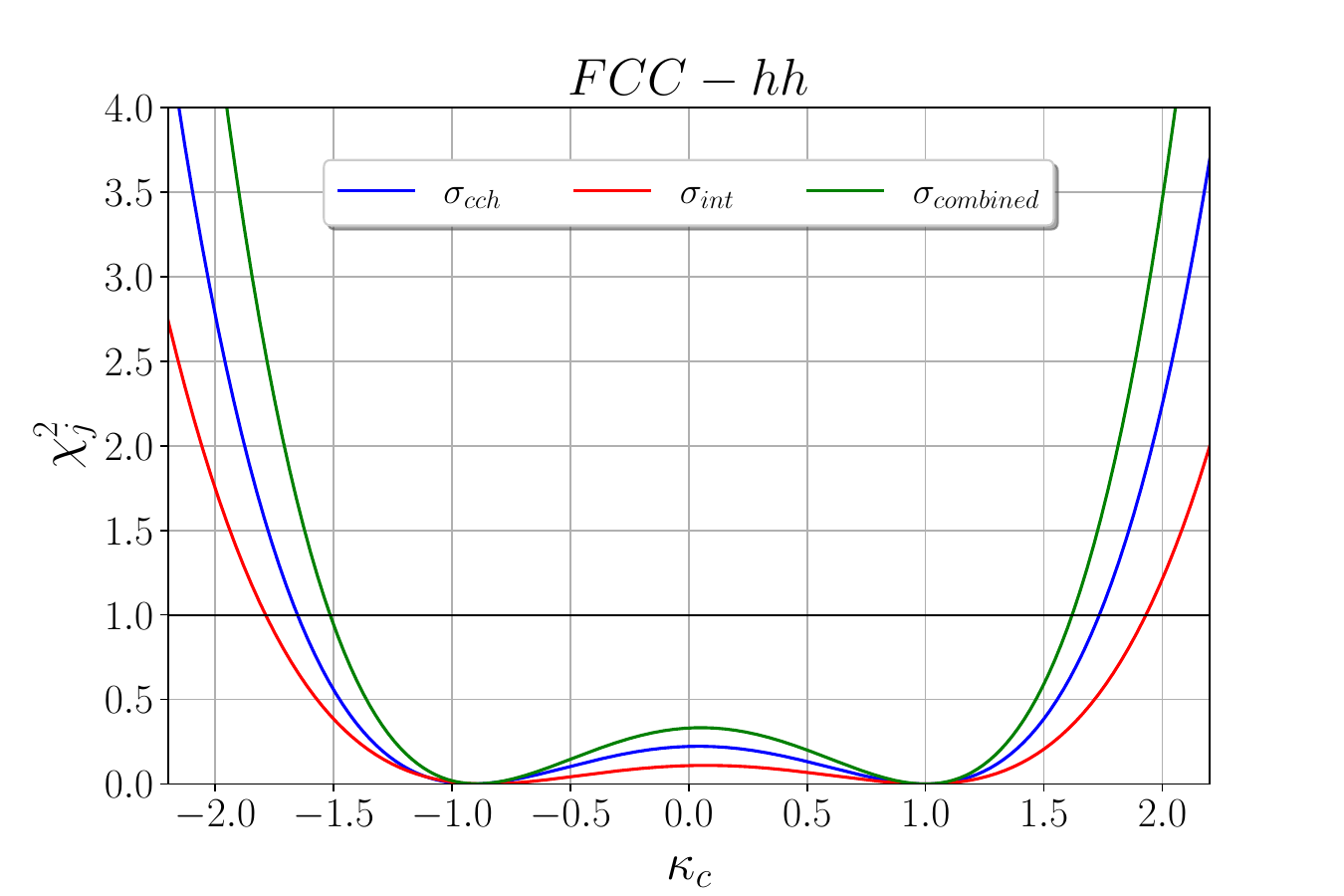}
        \caption{}
       
    \end{subfigure}
    \caption{The expected significance for a deviation from the SM as a function of $\kappa_c$ by combining significance from the two signal channels as defined in Eqn.~\ref{eq:signif1}. Panel (a) shows the expected significance at the HL-LHC (ATLAS+CMS, 6 ab$^{-1})$, and Panel (b) represents the significance at the FCC-hh (30 ab$^{-1})$.}
    \label{fig:cst_realkc}
\end{figure}

First we assume that charm Yukawa coupling remains real and the SM Lagrangian only changes in the charm Yukawa sector with a $\kappa_c$ rescaling:
\begin{equation}
     \mathcal{L} \supset -\kappa_c\frac{m_c}{v}\bar{c} ch .
\end{equation}
Here, $\kappa_c$ is a real number representing the deviation of the modified charm Yukawa from the SM value, and $v=246$ GeV is the SM vacuum expectation value. The Lagrangian corresponds to the SM when $\kappa_c=1$. Under our assumption, the correction for $cch$ is proportional to $\kappa^2_c$, while the correction for $int$ is proportional to $\kappa_c$. The correction for $ggh$ has a very minor dependence on $\kappa_c$ as well, whose numerical form is derived and included in Appendix \ref{app:kgkgamma} 
\footnote{For the Higgs decay, we assume a SM decay branching ratio to photons, where a correction from $\kappa_c$ affects the deviation at most at sub-percentage level, see \autoref{app:kgkgamma}.}. The respective $\chi$-square deviation from the SM by probing the three different signal channel can thus be calculated from the confusion matrix as:
\begin{equation}
     \chi^2_j = \frac{1}{\sum_i N_{ij}}\left(N_{4j}\kappa^2_c+N_{5j}\kappa_c(1.01-0.01\kappa_c)-N_{4j}-N_{5j}\right)^2.
\label{eq:signif1}
\end{equation}
Here $N_{ij}$ are the confusion matrix elements from the SM prediction.
When $j$ takes the value of 4 and 5, it represents the number of events in the $cch$ and $int$ columns of the confusion matrix, respectively. 
We plot the significance curves as a function of $\kappa_c$ in Fig.~\ref{fig:cst_realkc}. Notably, the dominant signal contribution is from $\kappa^2_c$, hence the mostly symmetric shape of the significance. Given limited sensitivity to the interference part, the probe gives bound region of $-5.6 <\kappa_c < 5.6$ (HL-LHC) and $ -1.51 < \kappa_c < 1.62$ (FCC-hh, slightly asymmetric) with 1$\sigma$ significance.

\subsection{Constraints on a CP-complex charm Yukawa coupling}

When including possible CP-phase in the charm Yukawa coupling, we consider a modified SM Lagrangian as follows:
\begin{equation}
     \mathcal{L} \supset -\frac{m_c}{v}|\kappa_c| \bar c (\cos{\alpha} + i\gamma_5 \sin{\alpha})ch .
\end{equation}
When $|\kappa_c|=1$ and $\alpha=0$, it returns to the SM. Here $|\kappa_c|$ denotes the rescaling of the overall magnitude of a CP-complex charm Yukawa coupling, and $\alpha$ the CP-phase. 
Alternatively, we can also parametrize in terms of the CP-real and imaginary part of the coupling with $\kappa_c = |\kappa_c|\cos{\alpha}$ and $\tilde{\kappa_c}= |\kappa_c|\sin{\alpha}$. The phase angle $\alpha$ does not affect the $cch$ channel contribution which is always proportional to $y_c^2$, but introduces a factor of $\cos{\alpha}$ in the $y_c$ dependent $int$ channel contribution. Such dependence arises from the relative phase between the modified $cch$-diagram and the $ggh$-diagram, with the latter mostly remain SM. Therefore, the chi-square deviation as a function of the charm Yukawa modification factors are now defined as follows:
\begin{equation}
\begin{aligned}
\chi^2_j &= \frac{1}{\sum_i N_{ij}} \left( N_{4j}|\kappa_c|^2 + (1.01|\kappa_c|\cos{\alpha} - 0.01|\kappa_c|^2 \right. \\
&\quad \left.- 0.001|\kappa_c|^2\sin^2{\alpha})N_{5j} - N_{4j} - N_{5j} \right)^2.
\end{aligned}
\end{equation}

Next we attempt to compare constraints on a CP-complex charm Yukawa coupling from this $ch$ study with other collider signals. 
As discussed in the introduction, the $Zh,Zh (h\to c\bar{c}$ channel offers a competitive probe of the charm Yukawa coupling with realistic data and collider simulation. At the HL-LHC assuming 3 ab$^{-1}$ integrated luminosity and 14 TeV center of mass energy, ATLAS prospects an upper limit of $\mu_{Zh (h\to c\bar{c})} < 6.3$~\cite{ATL-PHYS-PUB-2018-016} with 2$\sigma$. At FCC-hh assuming 6 ab$^{-1}$ integrated luminosity and 100 TeV center of mass energy, constraints of $|\kappa_c|<2.1$ at $2\sigma$ are expected from $Vh(c\bar c)$ analysis in Ref.~\cite{Perez:2015lra}. 
To compare with our study, we rescale the two bounds above to $|\kappa_c|<1.69$ for a 6 ab$^{-1}$ luminosity at the HL-LHC and $0.47<|\kappa_c|<1.33$ for a 30 ab$^{-1}$ luminosity at the FCC-hh at 1$\sigma$.  
Additionally, indirect constraints from the precisely-measured Higgs production and decay rate such as $gg\to h$ and $h\to \gamma\gamma$ have sensitivity in the charm Yukawa coupling through the quark loop contribution. The charm quark loop contributes at LO mostly through interfering with the dominant top loop (and $W$ loop for $\gamma\gamma h$) diagram. The relative CP-phase between diagrams thus affects the size of the interference directly.
We include in Appendix~\ref{app:kgkgamma} the dependence of $\kappa_{g,\gamma}$ and $\tilde\kappa_{g,\gamma}$ on $\kappa_{c}$ and $\tilde\kappa_{c}$ after integrating the quark loop contribution as function of the modified Yukawa coupling. The projected constraints on $\kappa_g$ and $\kappa_\gamma$ from the corresponding production and decay rate measurement can be directly mapped to bounds on the charm Yukawa coupling modification.
The projected 1$\sigma$ sensitivity at the HL-LHC (ATLAS + CMS combined, 6 ab$^{-1}$) for $\kappa_g$ ($0.8\%$), $\kappa_\gamma$ ($1.3\%$) are derived from the projected bounds on the production cross-section and decay branching ratios $\sigma_{ggh}$ ($1.6\%$), BR$(h\to\gamma\gamma)$ ($2.6\%$) from Figs.~28-29 of the HL-LHC projections study~\cite{Cepeda:2019klc}. The CP-phase contribution to the total cross section comes in through the dependence on $|\kappa_{g,\gamma}|^2 + |\tilde\kappa_{g,\gamma}|^2$. These bounds on $\kappa_{g,\gamma}$ are dominated by inclusive Higgs measurements assumed to have no additional Yukawa coupling or NP dependence other than the $\kappa_c$ considered. At FCC-hh, the expected sensitivity from a global $\kappa$-fit to $\kappa_g$, $\kappa_\gamma$ are about $0.49\%$, $0.29\%$ respectively, taken from Table.~3 of Ref.~\cite{deBlas:2019rxi} including experimental and theory uncertainties. 

\begin{figure}[htbp]
    \centering
    \begin{subfigure}{0.49\textwidth} % 调整子图宽度
        \centering
        \includegraphics[width=\linewidth]{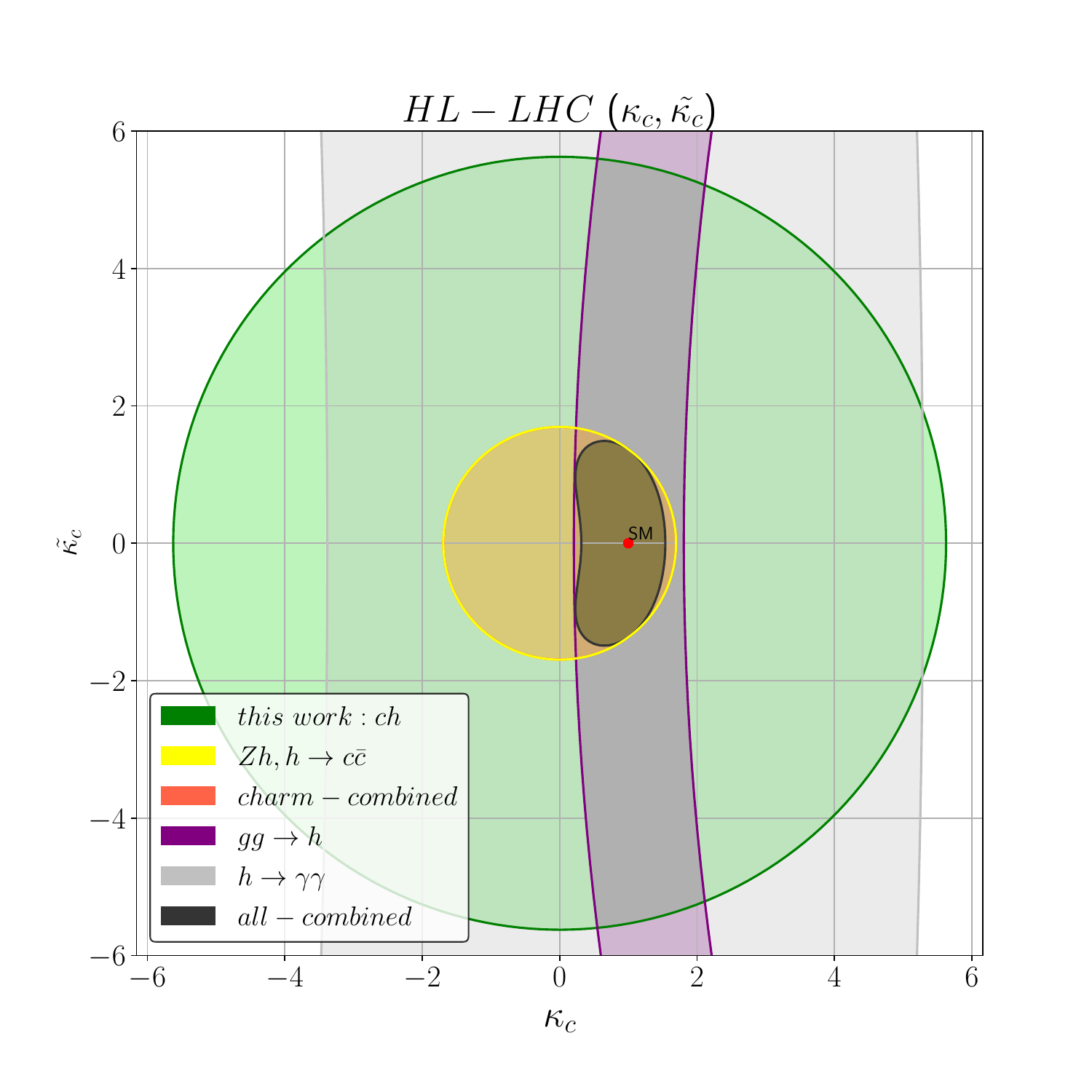}
        \caption{}
       
    \end{subfigure}
    \hfill
    \begin{subfigure}{0.49\textwidth} % 调整子图宽度
        \centering
        \includegraphics[width=\linewidth]{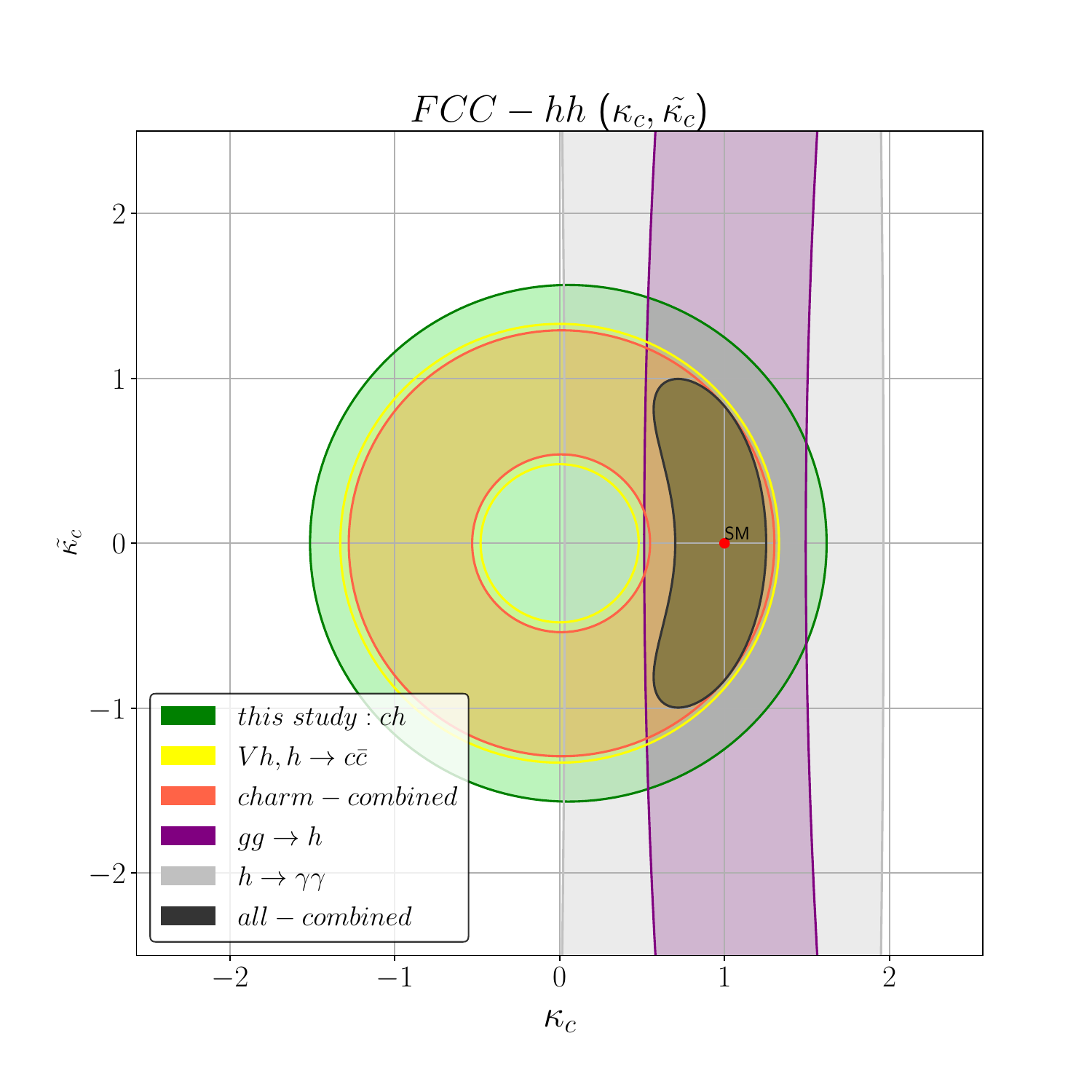}
        \caption{}
        
    \end{subfigure}
    \caption{(a) and (b) show the 1$\sigma$ sensitivity contours in the $(\kappa_c, \tilde{\kappa_c})$ space at the HL-LHC (ATLAS+CMS, 6 ab$^{-1}$) and the FCC-hh (30 ab$^{-1}$) respectively. All other parameters except for the charm Yukawa coupling are fixed to their SM values. The green region represents constraint from this $ch$ study, the yellow region from $Zh,~h \to c\bar{c}$, while the red region ``charm-combined" represents the bound from the two signals combined. Additionally, we show the purple and light gray contour regions which are indirect bounds from charm loop contribution to $gg\to h$ and $h\to \gamma\gamma$ inclusive Higgs production and decay measurements respectively.}
    \label{fig:collider-bounds}
\end{figure}

Taking all these collider signal into consideration, constraints on the complex charm Yukawa are shown in Fig.~\ref{fig:collider-bounds}, with HL-LHC on the left and the FCC-hh in the right panel. Reading from the plot, the most stringent constraint on the magnitude of the charm Yukawa comes from the measurement of the $Zh, h\to c\bar c$ (yellow region) process as expected. Being proportional to $|\kappa_{c}|^2 + |\tilde\kappa_{c}|^2$, it does not allow for the determination of the CP phase $\alpha_c$. It nevertheless probes the charm Yukawa contribution directly by identifying the charm jet in the final state as in our study. Bounds from our $ch$ analysis (green) is comparatively weak at the HL-LHC, but improved significantly at the FCC-hh. Since The large center of mass energy at the FCC significantly increased the cross section of the $ch$ process. In the plot, we also show the combined constraint achieved from our $ch$ study and $Zh,h\to c\bar c$ as ``charm-combined". Improvement to the bound by including the $ch$ study is about percent level, and slightly shift the bound from the center.
The indirect constraint from $gg\to h$ (purple region) production rate measurement offer sensitive probe to a large CP-space. There the charm loop contributes through interfering with the dominant top loop at percent level, which is reachable for the Higgs production through gluon fusion. Similar sensitivity are to be reached for $h\to \gamma\gamma$ (light gray region) decay rate measurement, whereas the charm loop generates a even minor contributions. The indirect constraints from $gg\to h$ offers a complementary probe of the CP-structure of the charm Yukawa coupling. 

Reading off from Fig.~\ref{fig:collider-bounds}, the study from the $ch$ channel alone give symmetric bounds of $|\kappa_c| < 5.6 $ at the HL-LHC and a slightly shifted bound of $ -1.50 < \kappa_c < 1.61 $, $ -1.57 < \tilde{\kappa_c} < 1.57$ at the FCC. 
While the bound from $Vh,~h\to c\bar c$ signal offers the best sensitivity, it is improved from $0.47<|\kappa_c|<1.33$ to $0.52<|\kappa_c|<1.28$ at the FCC, by about 12\%.
After combining with the Higgs inclusive production and decay rate measurement $gg\to h$ and $h\to\gamma\gamma$, a marginal constraint of $0.32<|\kappa_c|<1.69$, $ -77^\circ < \alpha < 77^\circ $ at the HL-LHC and $0.70<|\kappa_c|<1.29$, $ -55^\circ < \alpha < 55^\circ$ at the FCC-hh can be achieved.

\subsection{Bounds from EDM}
\begin{figure}[htbp]
    \centering     \includegraphics[width=0.5\linewidth, height=0.5\linewidth]{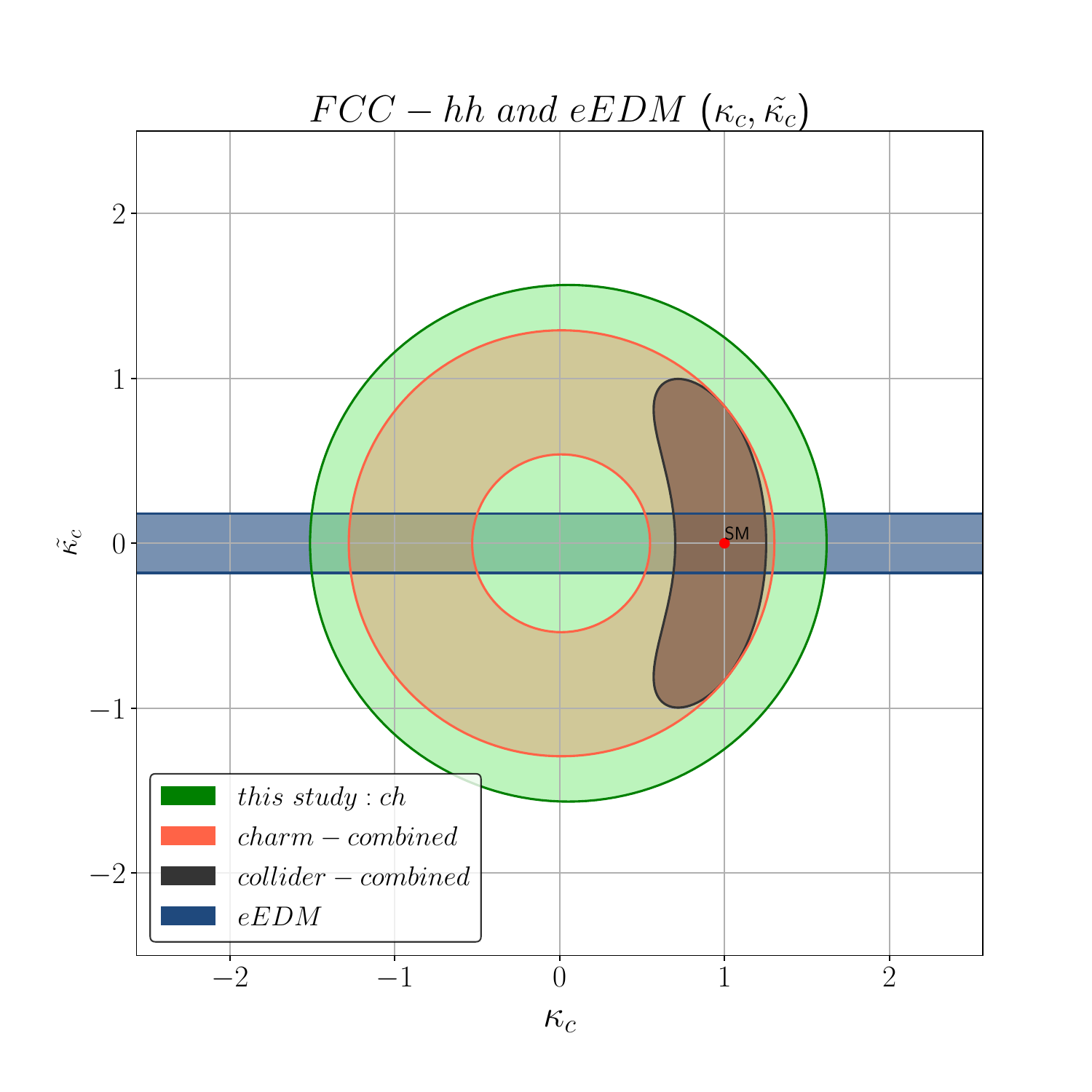}
        \caption{The 1$\sigma$ sensitivity contours are plotted in the $(\kappa_c,\tilde{\kappa_c})$ space while fixing all other parameters to their SM values for FCC-hh ( 30 ab$^{-1}$). The green and red contour correspond to interpretations of $ch$ and $charm-combined$ constraints, respectively. The dark purple contour represents the collider-combined constraints. The blue contour represents the constraints from the eEDM measurement. }
    \label{fig:edm-bounds}
\end{figure}

Electric dipole moment (EDM) measurements at low energies provide constraints on CP-odd components of the quark Yukawa couplings. Recent update from ACME~\cite{ACME:2018yjb} puts an upper bound on electron EDM (eEDM) at $\left| d_e \right| < 1.1 \times 10^{-29} e$ cm (at 90\% CL). Currently, the constraints on the charm Yukawa from eEDM are strongest compared to those from the neutron or other hadronic EDMs~\cite{Brod_2021}. The latest constraints on the charm-Yukawa from eEDM are $\tilde \kappa_c < 0.18$ (at $1\sigma$) from Equation 4.42 in Ref.~\cite{Brod:2023wsh}. The complementary constraints imposed by eEDM and the results of this work are shown in Fig.~\ref{fig:edm-bounds}. 
As expected, the EDM results sensitively constrain the CP-odd component of the charm Yukawa, while the collider results probe the overall magnitude more precisely.
It should be stressed though that, despite the seemingly stringent constraints from EDMs upon the CP-odd component, multiple flavor Yukawa couplings could contribute through the loop. Therefore, the probe of CP-odd term at collider processes where the charm flavor is exclusively identified such as in this study play an irreplaceable role. In addition, eEDM bounds rely on the assumption of a non-vanishing electron Yukawa coupling with the Higgs, which is hardly possible to test experimentally and may very well not hold under NP models. In this light, hadronic EDM measurements offers weaker yet complementary tests on the CP-odd component of the quark Yukawa couplings.

\section{Conclusion}
\label{sec:conclude}
In this work, we study the $ch~(h \to \gamma \gamma)$ signal and its probe to the charm Yukawa coupling. We perform full detector simulation and use BDT method with the calculation of Shapley values from an interpretable machine learning framework. We identify and classify the different types of signal and background contribution, along with an understanding of the kinematics spanned by the main collider observables. With the optimized classification results, we achieve an improved bounds on the charm Yukawa coupling as well as the CP-phase. By combining with collider signals as well as EDM measurements, we achieve a prospects for probing the CP-complex charm Yukawa coupling space at HL-LHC and FCC-hh. In summary, this work provides the following main conclusion: 
\begin{itemize}
\item At the HL-LHC, limited by statistics, we reach constraints of $|\kappa_c|<5.6$ at the $1\sigma$ level, and negligible constraints on the phase. After combining with the results from $Zh, h\to c \bar c$ and Higgs production rate ($g g \to h,~h \to \gamma \gamma$), we obtain $0.32<|\kappa_c|<1.69$, $ -77^\circ < \alpha < 77^\circ$ at 1$\sigma$ level.

\item At the FCC-hh, the significantly increased event numbers of the $ch$ effectively improved the precision of constraints. At the 1$\sigma$ level, an asymmetric constraint of $-1.51 < \kappa_c < 1.62$ is achieved for a real $\kappa_c$. This improves bound from $Vh,h\to c\bar c$ by about 12\%. Combing further with Higgs production and decay rate measurement we arrive a bound of $0.70<|\kappa_c|<1.29$, $ -55^\circ < \alpha < 55^\circ$. 

\item EDM measurements offer complementary probe to the charm Yukawa coupling as shown in Fig.~\ref{fig:edm-bounds}. Whereas the eEDM currently gives the most stringent bound on the CP-odd component of charm Yukawa coupling $\tilde\kappa_c$, the collider signal give best constraint on the overall magnitude. Notably, collider signals with exclusive charm jet identification along with the Higgs resonance signal such as the $h\to c\bar c$ and our study of the $ch$ provide more direct or indisputable probe into the charm Yukawa coupling space.

\item Machine learning methods, when combined with interpretable frameworks, provide valuable insights. We demonstrate in \autoref{sec:analysis} the shape distributions of different observables along with their corresponding absolute mean of Shapley values, a well defined importance measure in the BDT classification procedure. The contribution of qualitative shape difference and different correlation patterns among the complete set of observables can be better understood with the importance measure from the BDT machine learning process.
\end{itemize}

\appendix
\section{Effective \texorpdfstring{$\kappa_g$}{kappag} and \texorpdfstring{$\kappa_\gamma$}{kappagamma}}
\label{app:kgkgamma}
%%%%%%%%%%%%%%%%%%%

As described in the text, it is useful to list the numerical dependence of complex $ggh$ and $\gamma\gamma h$ effective coupling on a complex charm Yukawa and its modification. Following the procedure in Ref.~\cite{Brod:2013cka}, the couplings at leading order contribution are defined as, 
\begin{equation}
    \mathcal{L}_{\rm eff}\supset c_g \frac{\alpha_s}{12\pi}\frac{h}{v}G_{\mu\nu}^a G^{\mu\nu,a} + \tilde c_g \frac{\alpha_s}{8\pi}\frac{h}{v}G_{\mu\nu}^a \tilde G^{\mu\nu,a},
\end{equation}
\begin{equation}
    \mathcal{L}_{\rm eff}\supset c_\gamma \frac{\alpha}{\pi}\frac{h}{v}F_{\mu\nu} F^{\mu\nu} +  \tilde c_\gamma \frac{3\alpha}{2\pi}\frac{h}{v}F_{\mu\nu} F^{\mu\nu}.
\end{equation}
The CP-even and CP-odd couplings are defined respectively as,
\begin{equation}
    c_g = \sum_{f=t,b,c} \kappa_f A(\tau_f),~~~\tilde c_g = \sum_{f=t,b,c} \tilde\kappa_f B(\tau_f),
\end{equation}
with $\tau_f = 4m_f^2/m_h^2$ and
\begin{equation}
    A(\tau) = \frac{3\tau}{2}\left[ 1+ (1-\tau)\arctan^2\frac{1}{\sqrt{\tau -1}} \right],~~ B(\tau) = \tau\arctan^2\frac{1}{\sqrt{\tau -1}}.
\end{equation}
Hence, $\kappa_g$ and $\kappa_\gamma$, the rescaling of the effective $ggh$ and $\gamma\gamma h$ couplings, can be written as functions of $\kappa_c$ and $\tilde\kappa_c$ as,
\begin{equation}
    \kappa_g = \frac{c_g}{c_{g}^{\rm SM}} =  \frac{A(\tau_t) + A(\tau_b) + \kappa_c A(\tau_c)}
    { A(\tau_t) + A(\tau_b) + A(\tau_c)}
\end{equation}
\begin{equation}
    \tilde\kappa_g = \frac{3}{2}\frac{\tilde c_g}{c_{g}^{\rm SM}} = \frac{3}{2}\frac{\tilde\kappa_c B(\tau_c)}{ A(\tau_t)+ A(\tau_b) + A(\tau_c)}
\end{equation}
such that the modifier for inclusive gluon fusion rate is $\mu_g = |\kappa_g|^2 + |\tilde\kappa_g|^2$.

\begin{equation}
    \kappa_g \sim (-0.01 +0.01 i )\kappa_c + (1.01 - 0.01 i )
    \label{eq:kgkb}
\end{equation}
\begin{equation}
    \tilde\kappa_g \sim (-0.011 +0.01 i )\tilde\kappa_c 
\end{equation}
Similarly, the $h\gamma\gamma$ effective couplings can be expressed in terms of $\kappa_c$. In addition to the linear dependence on the Yukawa coupling, there is also a large constant term in the CP-even coupling from the leading $W$ and sub-leading top contributions at one loop in the SM. We omit here the well-known analytical forms and show the numerical dependence that matters for this analysis:
\begin{equation}
    \kappa_\gamma = (0.997 + 0.0024i) +  (0.003 - 0.0024i)\kappa_c 
\end{equation}
\begin{equation}
    \tilde\kappa_\gamma = (0.003 - 0.0024i)\tilde\kappa_c 
\end{equation}

\acknowledgments
We would like to thank Professor Kirill Melnikov giving explanation on available NLO calculations for the $cg\to ch$ process. ZQ is supported by the National Natural Science Foundation of China (1240050404) and the HZNU start-up fund.

%\normalem

\bibliographystyle{JHEP.bst}
\bibliography{biblio}
\end{document}